# Unconventional Orbital Magnetism in Graphene-based Fractional Chern Insulators

Jian Xie[1†], Zaizhe Zhang[1†], Xi Chen[1†], Yves H. Kwan[2†], Zihao Huo[1], Jonah Herzog-Arbeitman[3], Liangliang Guo[1], Kenji Watanabe[4], Takashi Taniguchi[5], Kaihui Liu[6], X.C. Xie[1,7,8], B. Andrei Bernevig[3,9,10]*, Zhi-Da Song[1,8,11]* and Xiaobo Lu[1,11]*

[1]International Center for Quantum Materials, School of Physics, Peking University, Beijing 100871, China
[2]Princeton Center for Theoretical Science, Princeton University, Princeton NJ 08544, USA
[3]Department of Physics, Princeton University, Princeton, NJ, 08544, USA
[4]Research Center for Electronic and Optical Materials, National Institute of Material Sciences, 1-1 Namiki, Tsukuba 305-0044, Japan
[5]Research Center for Materials Nanoarchitectonics, National Institute of Material Sciences, 1-1 Namiki, Tsukuba 305-0044, Japan
[6]State Key Laboratory for Mesoscopic Physics, Frontiers Science Centre for Nano-optoelectronics, School of Physics, Peking University, Beijing 100871, China
[7]Interdisciplinary Center for Theoretical Physics and Information Sciences, Fudan University, Shanghai 200433, China
[8]Hefei National Laboratory, Hefei 230088, China
[9]Donostia International Physics Center, P. Manuel de Lardizabal 4, 20018 Donostia-San Sebastian, Spain
[10]IKERBASQUE, Basque Foundation for Science, Bilbao, Spain
[11]Collaborative Innovation Center of Quantum Matter, Beijing 100871, China

†These authors contributed equally to this work.
*E-mail: xiaobolu@pku.edu.cn; songzd@pku.edu.cn; bernevig@princeton.edu

**Orbital magnetism in graphene originates from correlation-driven spontaneous valley symmetry breaking[1-7]. It can lead to various anomalous transport phenomena such as integer and fractional quantum anomalous Hall effects[8-11]. In general, the in-plane magnetic field $B_∥$ primarily couples to the spin degrees of freedom in graphene and has long been presumed to have a negligible effect on orbital magnetism due to the ultra-weak spin-orbit coupling[12-18]. In this work, we report multiple unconventional orbital magnetic phenomena that are highly sensitive to the $B_∥$ field in graphene/hBN superlattices hosting both integer and fractional Chern insulators (FCIs). We observed chirality-switching behaviors of the Chern insulator at moiré filling factor ν = 1 under a finite $B_∥$, demonstrating that both the $C = ±1$ states are permissible ground states at zero perpendicular magnetic field $B_⊥$. For the FCI at ν = 2/3, we observed topological phase transitions between two states characterized by Hall resistivity $ρ_{xy} = ±3h/2e^2$ under both $B_⊥$ and $B_∥$ fields. In-plane $B_∥$ field can effectively suppress the FCI state at zero $B_⊥$ field and enhance the FCI state with the opposite chirality, as resolved in Landau fan diagrams. Moreover, we observed rich phase transitions at 1 < ν < 2, accompanied by intervalley coherence and anomalous Hall effects (AHE) that can be triggered by sweeping either $B_⊥$ or $B_∥$. Our work has unveiled new properties of orbital magnetism, providing a new knob for engineering various AHE in graphene.**

Ferromagnetism, featuring long-range magnetic ordering and broken time-reversal symmetry, generically arises from nonzero spin magnetization or orbital magnetization[1-7,19-22]. In spin ferromagnets, spontaneous spin polarization is driven by exchange interactions which break spin-rotation invariance. Electrons can further acquire an orbital moment through spin-orbit coupling (SOC), leading to nonzero anomalous Hall conductivity. Unlike spin ferromagnets, orbital magnets simultaneously break time-reversal symmetry and develop spontaneous orbital moments without the need for SOC. Recently discovered orbital Chern insulators (CIs) in graphene-based superlattices, including its fractional version, have established the key role played by orbital magnetism in such topological phases[6-11,23-35]. Due to the out-of-plane nature of orbital moments, perpendicular magnetic fields are expected to manipulate these topological states in two-dimensional materials. In contrast, in-plane magnetic fields are in principle not expected to have a significant effect.

Rhombohedral multilayer graphene exhibits rich correlated physics at low excitations. Numerous competing symmetry-breaking states with nearly degenerate energies have been revealed under external tuning parameters such as electrical displacement field, magnetic field and carrier density[36-46]. However, studies of the interplay between in-plane magnetic fields and orbital-magnetism-based topological states including both integer and fractional Chern insulators (FCIs) remain scarce. Here we present the magneto-transport measurements in rhombohedral hexalayer graphene (RHG)/hBN superlattices hosting both CIs and FCIs. Surprisingly, we find an unexpected richness of topological phenomena, orchestrated by the interplay between the orbital magnetism and both in-plane and out-of-plane magnetic fields.

## $B_\parallel$-dependent CI and FCI

The schematic of the device configurations is shown in Fig. 1a, where the RHG sample is encapsulated with insulating hBN layers and two graphite gates. The RHG sample is further aligned with hBN on one side, resulting in twist angles of $\theta \approx 0.17°$ for device 1, 0.22° for device 2 and 0.25° for device 3 (Methods). Figure 1b,c displays the phase diagrams of longitudinal resistivity $\rho_{xx}$ and Hall resistivity $\rho_{xy}$ as a function of electric displacement field $D/\varepsilon_0$ and moiré filling factor $\nu = n/n_0$ (Methods) measured at different in-plane magnetic fields $B_\parallel$. When $B_\parallel = 0$, the CI state at $\nu = 1$ retains a single chirality at a small $B_\perp$ of 40 mT, exhibiting the $\rho_{xy}$ quantized to $h/e^2$ ($h$, the Planck constant; e > 0, the electron charge) and corresponding to $C = 1$ (Methods). However, an in-plane field of $B_\parallel = 2$ T splits the CI state into two parts as a function of filling factor with opposite chirality, while the corresponding $\rho_{xy}$ remains quantized to $-h/e^2$ and $h/e^2$, respectively. The chirality switching behavior can be more clearly revealed in the Landau fan diagrams under different $B_\parallel$ (Fig. 1d and e), where both $C = \pm 1$ states are well described by Streda's formula $\frac{\partial n}{\partial B_\perp} = \frac{Ce}{h}$. Notably, the $C = -1$ state requires a weak perpendicular magnetic field $B_\perp \approx 0.5$ T to be stabilized at zero $B_\parallel$, whereas above a finite critical $B_\parallel$ (for device 1, the critical value is $B_\parallel \approx 0.5$ T at $D/\varepsilon_0 = -0.53$ V nm$^{-1}$), the state with $B_\perp C < 0$ becomes one of the ground states even at $B_\perp = 0$, and coexists with the $B_\perp C > 0$ state in the Landau fan diagram. Landau fan diagrams under other values of fixed $B_\parallel$ exhibit similar chirality switching behaviors (Extended Data Fig. 1). Moreover, the $\rho_{xy}$ of the two states with opposite chirality near $\nu = 1$ remains quantized (Fig. 1f and g), with integer quantum anomalous Hall effect (IQAHE) observed simultaneously at fixed $\nu$ and $D$, but under different $B_\parallel$ (Fig. 1h). This indicates that the $|C| = 1$ CIs with different chirality can serve as ground states at $\nu = 1$ under finite $B_\parallel$, which effectively polarizes the spin along the

in-plane direction. Figure 1i shows the phase boundary of chirality reversal in the $B_\perp$-$B_\parallel$ parameter space, where the phase boundary approximates an elliptical shape. This suggests an effective interaction between the Chern number and $B_\parallel$, the latter of which is expected to mainly couple to spin degrees of freedom in systems with $C_{3z}$ symmetry such as RHG/hBN. The control of the Chern number by $B_\parallel$ has recently been reported in twisted bilayer–trilayer graphene[47], potentially with a similar origin. However, unlike in that case, the coexistence of CIs with both kinds of chirality be achieved here.

Tunable FCIs at $\nu = 2/3$ have been reported in rhombohedral graphene superlattices[8-11,30] soon after the initial discovery in the twisted MoTe$_2$ system[48-51]. However, the role of spin in these fractional states remain largely uncharted. In Fig. 2a-c, we present the symmetrized $\rho_{xx}$ versus $\nu$ and $B_\perp$ at various fixed $B_\parallel$. At $B_\parallel = 0.2$ T, the $\nu = 2/3$ state follows a trajectory described by Streda's formula, evolving up to $B_\perp \approx 0.9$ T, beyond which it transitions into a new phase characterized by an opposite sign of $\rho_{xy}$. Surprisingly, the critical value of $B_\perp$ triggering the phase transition becomes lower with increasing $B_\parallel$, resulting in the low-$B_\perp$ ground state being confined to a narrower range of $B_\perp$ (Fig. 2b,c). However, the new state at large $B_\perp$ with negative trajectory in $B_\perp$-$\nu$ parameter space does not completely take over the ground state at zero $B_\perp$, even when $B_\parallel$ increases to 2.8 T (Extended Data Fig. 2). This is further confirmed by the behavior of $\rho_{xy}$ (Fig. 2d-f), where the sign-reversal of $\rho_{xy}$ associated with the phase transition occurs at lower but non-vanishing $B_\perp$ with increasing $B_\parallel$. Notably, the sign-reversal of $\rho_{xy}$ reversal spans the entire filling range from $\nu = 0.4$ to $\nu = 1$, rather than being exclusive to the $\nu = 2/3$ FCI state. The competition between both states under $B_\parallel$ can be more intuitively revealed by the line cuts shown in Fig. 2g,h, which are extracted along the dashed lines in Fig. 2a-f. Unlike $\nu = 1$ state, where a sufficiently large $B_\parallel$ stabilizes the IQAHE with an opposite sign of $\rho_{xy}$, the FCI state at $\nu = 2/3$ does not exhibit new FQAHE with opposite chirality, even at large $B_\parallel$ which is sufficient to polarize spins in-plane. To better understand the role of spin as a function of $B_\parallel$, we further map the phase diagram of $\rho_{xy}$ in the $B_\perp$-$B_\parallel$ parameter space (Fig. 2i). Strikingly, the phase boundary corresponding to sign-reversal of $\rho_{xy}$ takes an approximately elliptical shape at small $B_\parallel$ (below 0.6 T) which is similar to the $C = \pm 1$ state at $\nu = 1$. However, the impact of $B_\parallel$ on the sign-reversal of $\rho_{xy}$ saturates at high $B_\parallel$, leaving a residual tail at low $B_\perp$ and high $B_\parallel$. Similar features of the topological phase transitions of FCI states have been observed in device 2 and 3 (Extended Data Fig. 3).

## Intervalley coherence and AHE at $1 < \nu < 2$

We now examine the $\nu > 1$ phase diagram of $\rho_{xx}$ shown in Fig. 3a. In addition to the CI at $\nu = 1$ and the topologically-trivial insulator at $\nu = 2$, there is a well-defined signature of Van Hove singularity (VHS) at $1 < \nu < 2$, emerging from a critical $D/\varepsilon_0 = -0.57$ V nm$^{-1}$. The VHS exhibits a resistance peak with weak insulating behavior (Extended Data Fig. 4). Figure 3b,c presents symmetrized $\rho_{xx}$ and antisymmetrized $\rho_{xy}$ as a function of $\nu$ and $B_\perp$ field measured at fixed $D/\varepsilon_0 = -0.58$ V nm$^{-1}$ (corresponding to the dashed line in Fig. 3a), exhibiting many distinct features. Firstly, the CIs with $C = \pm 1$ are dominant states at $\nu = 1$ and can coexist at finite $B_\perp$ field as discussed in the previous section. Secondly, the VHS truncates the Landau levels that would have extrapolated to $\nu = 2$ (Extended Data Fig. 5c,e) and confines a series of singly degenerate quantum Hall states within the associated dome in $D$-$\nu$ space (Fig. 3d and Extended Data Fig. 6). In contrast, the quantum Hall states above the VHS can be traced back to $\nu = 1$ at small $D/\varepsilon_0$ (Extended Data Fig.

5a). It is also noteworthy that position of the phase boundary does not follow the linear dependence on $B_\perp$ predicted by Streda's formula, and the slope is highly sensitive to $D$.

Moreover, the $\rho_{xy}$ shown in Fig. 3c exhibits a sharp sign reversal near $B_\perp = 0$ in the region above the VHS, indicating the possible presence of the anomalous Hall effect (AHE). In contrast, no signatures of the AHE are observed below the VHS, as confirmed by measuring magnetic hysteresis loops (Fig. 3g and Extended Data Fig. 7). Above the VHS, we propose that the ground state is a valley-imbalanced metal (VIM) as evidenced by the observation of AHE and singly degenerate Landau levels. In such a state, a whole band in the valley with $C = 1$ is fully filled and a band in the other valley with $C = -1$ is partially filled, yielding the VIM. Below the VHS, the observation of singly degenerate Landau levels and vanishing AHE suggests that the ground state is an intervalley-coherent metal (IVCM). The IVCM, where the electronic wavefunctions are coherent superpositions of the K and K' valleys[52-54], exhibits macroscopic phase coherence that can be visualized using scanning tunneling microscopy and spectroscopy[55-59]. We summarize the various phases in Fig. 3e, which shows $\rho_{xx}$ and $\rho_{xy}$ versus $\nu$ at $B_\perp = 1$ T. With increasing $\nu$, the system first transitions from the fully valley-polarized CI to the VIM, then into the IVCM, and finally returns to the VIM.

The phase boundary separating VIM and IVCM also depends on $B_\parallel$ (Extended Data Fig. 8). A finite $B_\parallel$ can notably enhance the VIM phase and suppress the IVCM phase at the same time. Strikingly, scanning $B_\parallel$ at $B_\perp = 0$ in the VIM phase, as well as the phase boundary, also reveals a notable AHE (Fig. 3f), manifested as a magnetic hysteresis window that is noticeably larger than that observed in $B_\perp$ scans. Taking $\nu = 1.09$ as an example, the magnetic hysteresis window for scanning $B_\parallel$ is $\Delta B_\parallel \approx 140$ mT, while the window for scanning $B_\perp$ is $\Delta B_\perp \approx 25$ mT (Fig. 3f,g). We have ruled out the possibility that the hysteresis observed in $B_\parallel$ is caused by the out-of-plane component (Extended Data Fig. 9). Moreover, similar hysteresis phenomena have also been observed in device 2 (Extended Data Fig. 10). Experimentally, Hall effects associated with $B_\parallel$ have been observed in certain antiferromagnetic materials and magnetic Weyl semimetals, though the AHE is absent in those systems[60-66]. The AHE triggered by $B_\parallel$ sweeping is reminiscent of systems with strong SOC or with proximate SOC[67] and has been reported in other graphene systems[68-70]. We also note that the sign of $B_\parallel$ can effectively control the sign of $\rho_{xy}$ at $1 < \nu < 2$, which is in sharp contrast with the behavior of the FCI at $\nu = 2/3$ and CI at $\nu = 1$.

## Possible mechanisms for the unconventional $B_\parallel$-dependance

Unlike the AHE at $1 < \nu < 2$, the coupling between $\rho_{xy}$ and $B_\parallel$ for the FCI at $\nu = 2/3$ and CI at $\nu = 1$ is invariant to the orientation of $B_\parallel$. As shown in Fig.1i and 2i, the sign of $\rho_{xy}$ is not locked to the sign of $B_\parallel$ even at sufficiently large values, which seems beyond the scope of in-plane orbital magnetization. To attempt to understand the topological phase transitions occurring at $\nu \leq 1$, we consider a phenomenological theory based on the competition between electron-electron interaction, intrinsic SOC, and orbital/spin magnetization. While the SOC strength is known to be only of the order of tens of μeV in graphene[13-18], it may have a significant effect on ground state properties due to the existence of numerous competing symmetry-breaking states with nearly degenerate energies in RHG/hBN. Electron-electron interaction favors states with larger pseudospin polarization. Without loss of generality, we assume $B_\perp > 0$. For $\nu \leq 1 - \frac{B_\perp A}{\Phi_0}$ (where $\Phi_0 = h/e$ and $A$ is the area of moiré unit cell), there are enough states to accommodate all electrons in any one of the pseudospin flavors. To minimize the interaction exchange energy, the electrons will therefore completely polarize into one of the flavors. The choice of flavor should further minimize the SOC and magnetization terms described by $H_{eff} = -B_\perp \overline{M_z} \tau_z + \mu_B B_\perp s_z + \mu_B B_\parallel s_x - \frac{\lambda_I}{2} \tau_z s_z$. Here $\mu_B = 5.79 \times 10^{-5}$ eV/T is the Bohr magneton, $\lambda_I \approx 40$ μeV is the intrinsic SOC strength, and $\tau_z = \pm 1$ denotes K and K' valley respectively (Methods). Since the two valleys are related by time reversal, the lowest moiré bands in the two valleys have opposite average orbital magnetization $\overline{M_z}\tau_z$, as well as opposite Chern number $C\tau_z$ with $|C| = 1$. From a phenomenological viewpoint, we must have either $0 < \overline{M_z} < \mu_B$, $C < 0$, $\lambda_I > 0$ or $-\mu_B < \overline{M_z} < 0$, $C > 0$, $\lambda_I < 0$ for this theory to align with experimentally observed Landau fan slopes (Methods). Let us assume $\lambda_I > 0$ for now, so that $0 < \overline{M_z} < \mu_B$, $C < 0$. (A similar analysis also applies to $\lambda_I < 0$.) In the absence of $B_\parallel$ and if $|\overline{M_z}B_\perp| > \lambda_I/2$, spin and valley magnetization dominate over SOC, making the ground state $|K, \downarrow>$ polarized. This necessitates $C < 0$ to be consistent with the observation of a CI whose Streda anomaly slopes towards $\nu \leq 1$. If $|\overline{M_z}B_\perp| < \lambda_I/2$, SOC dominates, locking the ground state polarization to either $|K', \downarrow>$ or $|K, \uparrow>$ (Fig.4a). Since $0 < \overline{M_z} < \mu_B$, $|K', \downarrow>$ is the ground state, which has opposite Hall conductance to $|K, \downarrow>$. A critical field strength $B_{\perp c} = \frac{|\lambda_I|}{2|\overline{M_z}|}$ separates the two limits when $B_\parallel = 0$. When a large $B_\parallel$ is applied, spin becomes in-plane polarized, suppressing $\langle s_z \rangle$ and thus intrinsic SOC. Therefore, $B_{\perp c}$ decreases as $B_\parallel$ increases. The calculated $B_{\perp c}$ dependence on $B_\parallel$ (Methods) is illustrated in Fig. 4b, with $|\lambda_I| \approx 40$ μeV obtained by fitting the experimental phase boundaries in the low $B_\parallel$ regime. Our estimation of $\lambda_I$ is of the same order of magnitude with previous theoretical[13-18] and experimental[39,59,71-76] estimations, which fall between 20 ~ 80 μeV. While our analysis is based on a phenomenological theory, this demonstrates that even such a weak SOC can potentially change the topological properties of graphene.

For $1 - \frac{B_\perp A}{\Phi_0} < \nu \leq 1 + \frac{B_\perp A}{\Phi_0}$, the Chern band in each spin sector in K' valley can hold more electrons than in the K valley due to the geometrical correction $\Omega \cdot B$ to the phase volume[77], where $\Omega$ is the Berry curvature (Methods). Thus, while electrons can be completely polarized to a spin sector in K' valley, each spin sector in K valley can only hold a portion of the electrons, making complete polarization to a spin sector in K valley impossible. For $\nu > 1 + \frac{B_\perp A}{\Phi_0}$, while no single flavor can hold all the electrons, maximum flavor polarization occurs when one spin sector in K' valley is fully occupied. Therefore, electron-electron interaction favors (partial) K' valley polarization for $\nu > 1 - \frac{B_\perp A}{\Phi_0}$, which competes with the effects of orbital/spin magnetization and SOC. The ground state valley polarization can be determined using a phenomenological Ginzburg-Landau theory (Methods), and is illustrated in the phase diagrams for $B_\parallel = 0$ and large $B_\parallel$ in Fig. 4c,d. Notably, while the $C = 1$ state always extrapolates to $B_\perp = 0$ regardless of $B_\parallel$, the $C = -1$ state can only extrapolate to $B_\perp = 0$ under strong $B_\parallel$, which agrees well with the experiment.

Given $\lambda_I > 0$, as obtained in DFT calculations (Methods) and consistent with the previous works[16,18], the above phenomenological theory requires a Chern number $C < 0$ in the K valley to align with experiments, which is inconsistent with Hartree-Fock calculations based on the effective model of rhombohedral graphene superlattices[78]. Specifically, the Hartree-Fock analysis yields $C = 1$ in the K valley (Methods). Furthermore, the Hartree-Fock calculations predict relatively large magnitudes of the orbital magnetization $|\overline{M_z}| > 5\mu_B$ (Methods), in contrast to the phenomenological theory. The origin of these discrepancies remains unresolved. This suggests that either the experimental observations originate from a mechanism beyond SOC, or the Hartree–Fock calculations based on existing continuum models of RHG/hBN may be inadequate, and we leave its resolution to future studies.

For the AHE at $1 < \nu < 2$, the experiment demonstrates that the sign of $\rho_{xy}$ can be controlled by the sign of $B_\parallel$, which could arise from different mechanisms. One mechanism is based on Rashba SOC which would break the in-plane spin rotation symmetry. However, Rashba SOC is unlikely to play a dominant role since it is negligible in RHG (Methods). Another mechanism is based on a significant in-plane orbital magnetization which may arise if $C_{3z}$ symmetry is strongly broken. Such an in-plane orbital magnetization would be opposite in the two valleys, and could enable a sufficient direct orbital coupling to $B_\parallel$.

Our experiments have revealed unconventional orbital magnetism over a wide range of moiré fillings in RHG/hBN. In particular, the qualitatively different dependence of both CI and FCI on $B_\parallel$ indicates that while the FCI may descend from the CI, its formation is governed by more intricate underlying physics. These aspects warrant further experimental and theoretical investigations. We noted a relevant study showing AHE tuned by $B_\parallel$ during the preparation of our manuscript[70].

## Acknowledgements

We thank Guangyu Zhang, Wei Yang, and Jingwei Dong for their experimental support. X.L. acknowledges support from the National Key R&D Program (Grant Nos. 2022YFA1403500 and 2024YFA1409002) and the National Natural Science Foundation of China (Grant Nos. 12274006 and 12141401). Z.-D.S. and X.C. were supported by National Natural Science Foundation of China (General Program No. 12274005), National Key Research and Development Program of China (No. 2021YFA1401900), and Innovation Program for Quantum Science and Technology (No. 2021ZD0302403). B.A.B. was supported by the Gordon and Betty Moore Foundation through Grant No. GBMF8685 towards the Princeton theory program, the Gordon and Betty Moore Foundation's EPiQS Initiative (Grant No. GBMF11070), the Office of Naval Research (ONR Grant No. N00014-20-1-2303), the Global Collaborative Network Grant at Princeton University, the Simons Investigator Grant No. 404513, the BSF Israel US foundation No. 2018226, the NSF-MERSEC (Grant No. MERSEC DMR 2011750), Simons Collaboration on New Frontiers in Superconductivity (SFI-MPS- NFS-00006741-01), and the Schmidt Foundation at the Princeton University, European Research Council (ERC) under the European Union's Horizon 2020 research and innovation program (Grant Agreement No. 101020833). X.X. acknowledges support from Innovation Program for Quantum Science and Technology (Grant No. 2021ZD0302400). K.W. and T.T. acknowledge support from the JSPS KAKENHI (Grant Nos. 21H05233 and 23H02052) and World Premier International Research Center Initiative (WPI), MEXT, Japan.

## Author Contributions

X.L. and J.X. conceived and designed the experiments; J.X. and Z.Z. fabricated the devices and performed the transport measurement with help from Z.H., L.G. and K.L.; J.X., X.C., Y.H.K., J.H.-A., X.X., B.A.B., Z.-D.S. and X.L. analyzed the data; Z.-D.S. and X.C. proposed the phenomenological theory and the SOC mechanism; Y.H.K., J.H.-A. and B.A.B. performed the

Hartree-Fock calculations; T.T. and K.W. provided the hBN crystals; J.X., X.C., Y.H.K., B.A.B., Z.-D.S. and X.L. wrote the paper with input from others.

## Competing interests

The authors declare no competing interests.

## Data Availability

All data supporting the findings of this study are available within the main text, figures and Supplementary Information, or from the corresponding authors upon request. Source data are provided with this paper.

## Code Availability

Codes that support the findings of this study are available upon request. Codes include scripts for data processing and theoretical modelling.

# Methods

### Device fabrication and transport measurement

For device 1 and 2, the method for device fabrication and basic characterization can be found in our previous work[9], where we detailed the fabrication process, graphene-layer number calibration, rhombohedral stacking identification, and alignment determination.

In this work, we have included the data from a new sample (device 3), where the RHG is aligned with the top-layer hBN, resulting in a moiré periodicity of $\lambda_m \approx 14.2$ nm (corresponding to the twist angle $\theta \approx 0.25°$). The topologically non-trivial states have been observed at the positive electric field where the electrons are pushed away from the moiré interface, which is consistent with the observations in device 1 and 2 (Extended Data Fig. 3).

The transport measurement methods and conditions are the same as the previous work[9]. The carrier density $n$ and electrical displacement field $D/\varepsilon_0$ were defined by the two graphite gate voltages, following $n = (C_{tg}V_{tg} + C_{bg}V_{bg})/e$ and $D/\varepsilon_0 = (C_{tg}V_{tg} - C_{bg}V_{bg})/(2\varepsilon_0)$, where $\varepsilon_0$ is the permittivity of vacuum, and $C_{tg}$ and $C_{bg}$ are the capacitances of the top and bottom gates per unit area, respectively. Moiré filling factor is given by $\nu = n/n_0$, where $n_0$ represents the carrier density $n$ required to fill one electron per moiré superlattice unit cell.

### Symmetrization and antisymmetrization

To eliminate geometric mixing, we symmetrized the longitudinal resistance $R_{xx}$ and antisymmetrized Hall resistance $R_{xy}$ separately during the data processing as follows:

$$R_{xx}^{\text{sym}}(B_\perp) = \frac{R_{xx}^{\text{raw}}(B_\perp) + R_{xx}^{\text{raw}}(-B_\perp)}{2},$$

$$R_{xy}^{\text{antisym}}(B_\perp) = \frac{R_{xy}^{\text{raw}}(B_\perp) - R_{xy}^{\text{raw}}(-B_\perp)}{2},$$

where the superscript 'raw' represents the unprocessed raw data.

We use resistivity $\rho$ instead of resistance $R$ for characterization

$$\rho_{xx} = R_{xx}\frac{w}{L}, \quad \rho_{xy} = R_{xy},$$

where the width/length ratio of the channel $w/L$ is 1/3 for device 1 and 1/2 for device 2.

## Theoretical calculation model

### Conventions and definitions of the model

For simplicity, the conventions of spin, valley and sublattice degrees of freedom, as well as the definition of Chern number in this work are the same as those in the previous works[78,79]. Specifically, the two in-plane lattice vectors are $\boldsymbol{a}_{1,2} = (\frac{a}{2}, \pm\frac{\sqrt{3}}{2}a)$, where $a$ is the graphene lattice constant. A and B orbitals of the same layer in one unit cell satisfy $\boldsymbol{r}_B - \boldsymbol{r}_A = (0, \frac{\sqrt{3}}{3}a)$. There are 12 orbitals $A_n, B_n$ in each unit cell, with $1 \leq n \leq 6$. We assume the +z direction points from the 1st layer to the 6th layer. $A_n$ lies directly above $B_{n-1}$ for $n \geq 2$, while $A_1, B_1$ is above the aligned hBN. Therefore, the lowest conduction (valence) band is manly contributed by $B_6(A_1)$ for $D < 0$. The K valley ($\tau_z = 1$) is defined as $\mathbf{K} = (\frac{4\pi}{3a}, 0)$, and $\mathbf{K}' = -\mathbf{K}$. The Berry connection and Berry curvature are defined as $A_i(\boldsymbol{k}) = i\langle u(\boldsymbol{k})|\partial_{k_i}|u(\boldsymbol{k})\rangle$ and $\Omega_z(\boldsymbol{k}) = \partial_{k_x}A_y(\boldsymbol{k}) - \partial_{k_y}A_x(\boldsymbol{k})$, where $\boldsymbol{k}$ is the Bloch momentum. The Chern number $C = \frac{1}{2\pi}\iint_{\boldsymbol{k}\in BZ}\Omega_z(\boldsymbol{k})d\boldsymbol{k}$. Under this convention[77], the Streda anomaly takes the form $\frac{\partial n}{\partial B_\perp} = \frac{Ce}{h}$. Thus, a positive (negative) Landau fan slope reveals a positive (negative) Chern number.

By definition, the Chern number is an integer. Accordingly, $C$ denotes the Chern number for CIs, while for FCIs, $C$ refers to the anomalous Hall resistivity quantization number, equivalent to the slope given by the Streda's formula.

### Effective spin-orbit coupling

The SOC of graphene under external electric field $D$ consists of intrinsic (KM) and extrinsic (BR) terms[12]: $H_{SOC} = H_{KM} + H_{BR} = \frac{\lambda_I}{2}\tau_z\sigma_z s_z + \lambda_{BR}(\tau_z\sigma_x s_y - \sigma_y s_x)$, where the extrinsic SOC strength $\lambda_{BR} \propto D$ is estimated to be $\approx 5$ µeV for $D/\varepsilon_0 = 1$ V nm$^{-1}$ (ref.16,17). With a negative displacement field, the lowest conduction (valence) band of RHG is mainly contributed by the B (A) sublattice of the top (bottom) layer. Since $H_{BR}$ is only nonzero between different sublattices in the same layer, it can only couple the lowest conduction band to higher energy bands, which are energetically separated from the lowest two bands by $\Delta E \approx 300$ meV. The effect of $H_{BR}$ on the lowest conduction band can thus be estimated with second-order perturbation theory, with energy scale $\frac{\lambda_{BR}^2}{2\Delta E} \approx 10^{-4}$ µeV, which is neglectable. Since $\langle \sigma_z \rangle \approx -1$ for the lowest conduction band for $D < 0$, the effective intrinsic SOC for $\nu > 0$ is $-\frac{\lambda_I}{2}\tau_z s_z$, coupling the spin and valley degrees of freedom.

The parameter $\lambda_I$ can be obtained from first-principles calculations. Consider a RHG system (moiré is unnecessary here) with a negative displacement field $D/\varepsilon_0 < 0$. Even for a small $|D/\varepsilon_0| \approx 0.1$ V nm$^{-1}$, the lowest conduction and valence bands are separated by an energy scale much larger than $\lambda_I$. Since the extrinsic SOC is neglectable for the lowest bands, $\lambda_I$ can be extracted from the spin splitting of the lowest conduction (valence) band at K point. We perform a DFT calculation within the local density approximation (LDA) using VASP (Vienna *ab initio* simulation package)[80], and find that if we use the "C_GW_new" pseudopotential for the carbon atom (with maximal energy cutoff 413.992 eV and includes three *s*, two *p*, two *d*, and one *f* basis vectors for the AE/PS partial waves), the spin up state at K point is 42 µeV lower than the spin down state, indicating $\lambda_I = 42$ µeV. The energy convergence criteria are set to $10^{-3}$ µeV due to the small value of $\lambda_I$, whose convergence with regard to the k-mesh and energy cutoff is carefully checked. However, the exact value of $\lambda_I$ depends on the choice of pseudopotential. For example, using the "C_h_GW" pseudopotential (with maximal energy cutoff 742.464 eV and includes three *s*, three *p*, and two *d* basis vectors for the AE/PS partial waves) would result in $\lambda_I = 24$ µeV. We note that for such a small energy scale, the result based on pseudopotentials may not be quantitatively accurate. Nevertheless, the calculation is expected to give the correct sign and order of magnitude for $\lambda_I$. Previous theoretical studies also suggest $\lambda_I > 0$ using symmetry analysis, with $|\lambda_I| \approx 24$ µeV obtained from all-electron scheme calculations[16,18]. We will assume $\lambda_I > 0$ from now on.

**Phenomenological theory for chirality switching at v ≤ 1**

Consider a simplified Hamiltonian with only a single effective moiré band per spin and valley:

$$\hat{H} = \sum_{\tau,s,\boldsymbol{k}} \epsilon_{\tau,s}(\boldsymbol{k}) c^\dagger_{\boldsymbol{k},\tau,s} c_{\boldsymbol{k},\tau,s} + \frac{1}{2N} \sum_{\tau,\tau',s,s',\boldsymbol{k},\boldsymbol{p},\boldsymbol{q}} V(\boldsymbol{q}) c^\dagger_{\boldsymbol{k}+\boldsymbol{q},\tau,s} c^\dagger_{\boldsymbol{p}-\boldsymbol{q},\tau',s'} c_{\boldsymbol{p},\tau',s'} c_{\boldsymbol{k},\tau,s}$$

where $\tau = \pm$ denotes K (K') valley, $s = \pm$ denotes spin up (down), $N$ is the number of moiré unit cells and $c_{\boldsymbol{k},\tau,s}$ ($c^\dagger_{\boldsymbol{k},\tau,s}$) is the electron annihilation (creation) operator. Note that $\hat{H}$ neglects aspects such as the form factors of the moiré band. Within the Hartree-Fock approximation, the total energy is

$$E_{tot} = E_{single} + E_{Hartree} + E_{Fock}$$

where

$$E_{single} = \sum_{\tau,s,\boldsymbol{k}} \epsilon_{\tau,s}(\boldsymbol{k}) n_{\tau,s}(\boldsymbol{k})$$

$$E_{Hartree} = \frac{V(\boldsymbol{0})}{2N} \sum_{\tau,\tau',s,s',\boldsymbol{k},\boldsymbol{p},\boldsymbol{q}} n_{\tau,s}(\boldsymbol{k}) n_{\tau',s'}(\boldsymbol{k})$$

$$E_{Fock} = -\frac{1}{2N} \sum_{\tau,s,\boldsymbol{k},\boldsymbol{p}} V(\boldsymbol{p}-\boldsymbol{k}) n_{\tau,s}(\boldsymbol{k}) n_{\tau',s'}(\boldsymbol{p})$$

and $0 \le n_{\tau,s}(\boldsymbol{k}) \le 1$ is the occupation. By replacing the sums with integrals $\Sigma_{\boldsymbol{k}} \to AN \int \frac{d^2k}{(2\pi)^2}$, where $A$ is the area of the moiré unit cell, the total energy per moiré unit cell can be expressed as

$$\frac{E_{tot}}{N} = \frac{E_{single}}{N} + \frac{E_{Hartree}}{N} + \frac{E_{Fock}}{N}$$
$$= \sum_{\tau,s} A \int \frac{d^2\boldsymbol{k}}{(2\pi)^2} \epsilon_{\tau,s}(\boldsymbol{k}) n_{\tau,s}(\boldsymbol{k})$$
$$+ \frac{A^2}{2} V(\boldsymbol{0}) \int \frac{d^2\boldsymbol{k} d^2\boldsymbol{p}}{(2\pi)^4} \sum_{\tau,\tau',s,s'} n_{\tau,s}(\boldsymbol{k}) n_{\tau',s'}(\boldsymbol{p})$$
$$- \frac{A^2}{2} \int \frac{d^2\boldsymbol{k} d^2\boldsymbol{p}}{(2\pi)^4} \sum_{\tau,s} V(\boldsymbol{p}-\boldsymbol{k}) n_{\tau,s}(\boldsymbol{k}) n_{\tau,s}(\boldsymbol{p})$$

Let us first consider $B_\parallel = 0$. $B_\perp$ enters the Hamiltonian through two effects: (i) the spin and orbital magnetization $\epsilon_{\tau,s}(\boldsymbol{k}) \to \epsilon_{\tau,s}(\boldsymbol{k}) - \tau M_z(\boldsymbol{k}) B_\perp + s\mu_B B_\perp$, and (ii) the geometric correction to the phase volume $d^2\boldsymbol{k} \to d^2\boldsymbol{k}(1 + \frac{e}{\hbar}\tau\Omega_z(\boldsymbol{k})B_\perp)$, where $M_z(\boldsymbol{k})$ and $\Omega_z(\boldsymbol{k})$ satisfy $A\int \frac{d^2\boldsymbol{k}}{(2\pi)^2} M_z(\boldsymbol{k}) = \overline{M_z}$ and $\int d^2\boldsymbol{k}\, \Omega_z(\boldsymbol{k}) = 2\pi C$. For simplicity, we ignore the $\boldsymbol{k}$ dependence of $n_{\tau,s}(\boldsymbol{k})$, and simplify the zero-field dispersion as $\epsilon_{\tau,s}(\boldsymbol{k}) \to \epsilon_0 - \lambda_I \tau s/2$, and the interaction potential as $V(\boldsymbol{0}) \to U, V(\boldsymbol{k} \ne \boldsymbol{0}) \to V$. To the linear order of $B_\perp$, the three parts of the total energy are

$$\frac{E_{single}}{N} = \sum_{\tau,s} n_{\tau,s}[(\epsilon_0 - \frac{1}{2}\lambda_I \tau s)(1 + \tau\frac{CA}{\Phi_0}B_\perp) - \tau \overline{M_z} B_\perp + s\mu_B B_\perp]$$

$$\frac{E_{Hartree}}{N} = \frac{U}{2} \sum_{\tau,\tau',s,s'} n_{\tau,s} n_{\tau',s'}(1 + (\tau+\tau')\frac{CA}{\Phi_0}B_\perp)$$

$$\frac{E_{Fock}}{N} = -\frac{V}{2} \sum_{\tau,s} n_{\tau,s}^2 (1 + 2\tau \frac{CA}{\Phi_0} B_\perp)$$

where $\Phi_0 = h/e$ is the flux quantum. We now define the filling in each flavor $\nu_{\tau,s} = n_{\tau,s}(1 + \tau\frac{CA}{\Phi_0}B_\perp)$, which is different from $n_{\tau,s}$ due to the nontrivial topology (Streda anomaly) of the lowest moiré band. We will assume $|C| = 1$ from now on. To the linear order of $B_\perp$, the three parts of the total energy can also be expressed as functions of $\nu_{\tau,s}$, which sum to a fixed total filling $\nu = \sum_{\tau,s} \nu_{\tau,s}$ and can be viewed as order parameters satisfying additional constraints $0 \le \nu_{\tau,s} \le (1 + \tau\frac{CA}{\Phi_0}B_\perp)$.

$$\frac{E_{single}}{N} = \sum_{\tau,s} \nu_{\tau,s}(\epsilon_0 - \frac{1}{2}\lambda_I \tau s - \tau \overline{M_z} B_\perp + s\mu_B B_\perp)$$

$$\frac{E_{Hartree}}{N} = \frac{1}{2} U \nu^2$$

$$\frac{E_{Fock}}{N} = -\frac{1}{2}V\sum_{\tau,s}\nu_{\tau,s}^2$$

where $E_{Hartree}$ is independent of the order parameters for fixed $\nu$, and $E_{Fock}$ favors states where electrons are concentrated to the same pseudospin flavor(s). Let us first consider $\nu \leq 1 - AB_\perp/\Phi_0$. $E_{Fock}$ can be minimized by polarizing all electrons to any of the four pseudospin flavors. The choice of flavor should further minimize $E_{single}$, which is described by the effective Hamiltonian

$$H_{eff} = -\overline{M_z}B_\perp \tau_z + \mu_B B_\perp \tau_z - \frac{\lambda_I}{2}\tau_z s_z$$

where both $\tau_z, s_z$ are good quantum numbers. Assuming $\lambda_I > 0$, if $\overline{M_z} < 0$ ($\overline{M_z} > \mu_B$), the ground state always has $\tau_z = -1$ ($\tau_z = 1$) regardless of $B_\perp$, in contradiction to the existence of a chirality reversal at some critical $B_{\perp c}$. Therefore, we must have $0 < \overline{M_z} < \mu_B$ for this theory to exhibit a valley-flip transition, and hence a sign-reversal of $\rho_{xy}$ for some critical $B_\perp$ as observed in experiments. In this case, the ground state has $\{\tau_z, s_z\} = \{-1, -1\}$ for $0 < B_\perp < \frac{\lambda_I}{2\overline{M_z}}$, and $\{\tau_z, s_z\} = \{1, -1\}$ for $B_\perp > \frac{\lambda_I}{2\overline{M_z}}$. To explain the positive (negative) Landau fan slope at $B_\perp > B_{\perp c}$ ($B_\perp < B_{\perp c}$) for the $\nu = 2/3$ state, we must have $C = -1$.

For $1 - \frac{AB_\perp}{\Phi_0} < \nu \leq 1 + \frac{AB_\perp}{\Phi_0}$, the lowest moiré band in each spin sector in K' valley with +1 Chern number can hold all electrons, while the lowest band in each spin sector in K valley cannot. For $B_\perp < \frac{\lambda_I}{2\overline{M_z}}$, the $\nu_{\tau,s} = \delta_{\tau,-1}\delta_{s,-1}\nu$ state (completely K' polarized) simultaneously minimizes $E_{single}$ and $E_{Fock}$, and is thus the ground state. For $B_\perp > \frac{\lambda_I}{2\overline{M_z}}$, the "ground state" flavor of $H_{eff}$ is $\{\tau, s\} = \{1, -1\}$, and the "first excited" flavor is $\{\tau, s\} = \{-1, -1\}$, which have enough states to accommodate all electrons. There will be no further energy reduction for the electrons to occupy "higher excited" flavors, so we can assume $\nu_{\tau,s} = \delta_{\tau,1}\delta_{s,-1}\nu_a + \delta_{\tau,-1}\delta_{s,-1}(\nu - \nu_a)$, where $0 \leq \nu_a = \nu_{+-} \leq 1 - \frac{AB_\perp}{\Phi_0}$ is the only independent order parameter. The total energy is

$$\frac{E_{tot}[\nu_a]}{N} = \left(-\mu_B B_\perp + \overline{M_z}B_\perp - \frac{\lambda_I}{2}\right)\nu - \frac{V}{2}\nu^2 + (\lambda_I - 2\overline{M_z}B_\perp + V\nu)\nu_a - V\nu_a^2$$

$\nu_a = 0$ signals the completely K' polarized state, and $\nu_a = 1 - \frac{AB_\perp}{\Phi_0}$ is the partially K polarized state. The chirality switching happens at $E_{tot}[0]/N = E_{tot}[1 - AB_\perp/\Phi_0]/N \Rightarrow \nu_c = 1 - \frac{\lambda_I}{V} - (\frac{A}{\Phi_0} - \frac{2\overline{M_z}}{V})B_\perp$. A reasonable estimation of $V \approx 10~meV$ results in $\frac{AV}{\Phi_0} \sim 8\mu_B > 2\mu_B > 2\overline{M_z}$, and $\frac{\lambda_I}{V} \sim 4 \times 10^{-3} \ll 1$. For $\nu > 1 + \frac{AB_\perp}{\Phi_0}$, the electrons must occupy at least two different flavors. As long as $|\nu - (1 + \frac{AB_\perp}{\Phi_0})|$ is sufficiently small, the ground state will be partially K' polarized regardless of $B_\perp$. The phase diagram of ground state valley polarization is summarized in Fig.4c. Specifically, the $C$ = -1 Landau fan crosses the magnetic phase boundary at $B_{\perp c} > 0$, and therefore does not extrapolate to $B_{\perp c} = 0$ when $B_\parallel = 0$, in agreement with experiment.

We now consider the effect of $B_\parallel$. In the $B_\parallel \to \infty$ limit, the spin is completely in-plane polarized with $\langle s_z \rangle = 0$, so the spin index and any term with $s_z$ in the above analysis should be omitted. Paralleling the derivation for $B_\parallel = 0$, it is straightforward to find that the ground state is completely K polarized, partially K polarized, completely K' polarized and partially K' polarized for $\nu \leq 1 - \frac{AB_\perp}{\Phi_0}$, $1 - \frac{AB_\perp}{\Phi_0} < \nu \leq \nu_c$, $\nu_c < \nu \leq 1 + \frac{AB_\perp}{\Phi_0}$ and $\nu > 1 + \frac{AB_\perp}{\Phi_0}$ respectively, where $\nu_c = 1 - (\frac{A}{\Phi_0} - \frac{2\overline{M_z}}{V})B_\perp$, as illustrated in Fig.4d. Specifically, there is no longer chirality switching at finite $B_{\perp c}$, and both $C = \pm 1$ Landau fans extrapolate to $B_{\perp c} = 0$ in this limit.

When $B_\parallel$ is finite and $\nu \leq 1 - \frac{AB_\perp}{\Phi_0}$, the electrons should still polarize to one flavor to minimize $E_{Fock}$. However, the effective Hamiltonian for $E_{single}$ is now

$$H_{\text{eff}} = -B_\perp \overline{M_z}\, \tau_z + \mu_B B_\perp s_z + \mu_B B_\parallel s_x - \frac{\lambda_I}{2} \tau_z s_z$$

with eigenvalues $E_{\tau_z,\pm} = -B_\perp \overline{M_z}\, \tau_z \pm \sqrt{\left(\mu_B B_\perp - \frac{\lambda_I}{2}\tau_z\right)^2 + \mu_B B_\parallel^2}$. Only the $E_{\tau_z,-}$ states are possible ground states. The magnetic phase boundary satisfies $E_{+,-} = E_{-,-}$, or

$$\left(\frac{B_\perp}{a}\right)^2 + \left(\frac{B_\parallel}{b}\right)^2 = 1$$

where $a = \frac{\lambda_I}{2|\overline{M_z}|}, b = \frac{\lambda_I}{2}\sqrt{(\overline{M_z})^{-2} - \mu_B^{-2}}$. The phase boundary is a semi-ellipse whose major/minor axes lengths depend on $\overline{M_z}$ and thus the displacement field. The difference of the two semi-ellipses in Fig. 4b is due to different $\overline{M_z}$, which originate from the different fillings and displacement field strength. Treating $\overline{M_z}$ as an unknown parameter, the experimental phase boundaries of both fillings in the low $B_\parallel$ regime can be well fitted by setting the SOC strength to $\lambda_I = 40$ μeV. The fitted $\overline{M_z}$ are $0.36\mu_B$ and $0.58\mu_B$ for $\nu = 2/3$ and $\nu = 0.94$, respectively. The displacement field is fixed at -0.74 V nm$^{-1}$ for $\nu = 0.94$ and -0.53 V nm$^{-1}$ for $\nu = 2/3$.

**Self-consistent Hartree-Fock calculations of RHG/hBN**

**A. Continuum model**

We first describe the continuum model for rhombohedral n-layer graphene (RnG) in the absence of a moiré potential. The low-energy physics can be expanded around the two valleys $\eta K$ of graphene, where $\eta = \pm$ indexes the valleys. The Hamiltonian for valley $+K$ is[78]

$$H_K(\boldsymbol{p}) = \begin{pmatrix} v_F \boldsymbol{p} \cdot \boldsymbol{\sigma} & t^\dagger(\boldsymbol{p}) & & t'^\dagger \\ t(\boldsymbol{p}) & \ddots & \ddots & t'^\dagger \\ t' & \ddots & v_F \boldsymbol{p} \cdot \boldsymbol{\sigma} & t^\dagger(\boldsymbol{p}) \\ t' & t(\boldsymbol{p}) & & v_F \boldsymbol{p} \cdot \boldsymbol{\sigma} \end{pmatrix} + H_{ISP} + H_D$$

$$t(\boldsymbol{p}) = -\begin{pmatrix} v_4 p_+ & -t_1 \\ v_3 p_- & v_4 p_+ \end{pmatrix}, \quad t' = \begin{pmatrix} 0 & 0 \\ t_2 & 0 \end{pmatrix}$$

$$[H_{ISP}]_{l\sigma,l'\sigma'} = V_{ISP} \left| l - \frac{n-1}{2} \right| \delta_{l,l'} \delta_{\sigma,\sigma'}$$

$$[H_D]_{l\sigma,l'\sigma'} = V \left( l - \frac{n-1}{2} \right) \delta_{l,l'} \delta_{\sigma,\sigma'},$$

where $\boldsymbol{p}$ is measured from $\boldsymbol{K}_G = \left(\frac{4\pi}{3a_G}, 0\right)$, $p_\pm = p_x \pm i p_y$, and $a_G = 2.46$ Å is the graphene lattice constant. $H_K(\boldsymbol{p})$ is a $2n \times 2n$ matrix in layer ($l = 0, \ldots, n-1$) and sublattice ($\sigma = A, B$) space, and is ordered according to $(0, A), (0, B), (1, A), \ldots, (n-1, B)$, where $(l, \sigma)$ indexes the layer $l$ and sublattice $\sigma$ degree of freedom. $\boldsymbol{\sigma} = (\sigma_x, \sigma_y)$ are Pauli matrices in sublattice subspace. An externally applied displacement field is implemented in $H_D$ as a linearly varying layer potential of amplitude $V$. We consider the RnG parameters $v_F = 660.4$ meV nm, $t_1 = 380$ meV, $t_2 = -10.5$ meV, $v_3 = 61.8$ meV nm, $v_4 = 30$ meV nm, $V_{ISP} = 0$ meV (ref. 81).

The RnG/hBN superlattice is constructed by aligning the hBN substrate adjacent to the bottom layer ($l = 0$) of graphene with twist angle $\theta$. The resulting moiré pattern is characterized by the moiré wavevector

$$\boldsymbol{q}_1 = \frac{4\pi}{3a_G} \left( 1 - \frac{1}{1 + \epsilon_{lat}} R(-\theta) \right) \hat{x},$$

where $R(\theta)$ is a counter-clockwise rotation by $\theta$, and $\epsilon_{lat} = (a_{hBN} - a_G)/a_G \simeq 0.017$ parameterizes the lattice mismatch between hBN and graphene. The basis moiré reciprocal lattice vectors (RLVs) are $\boldsymbol{b}_1 = \boldsymbol{q}_2 - \boldsymbol{q}_3, \boldsymbol{b}_2 = \boldsymbol{q}_3 - \boldsymbol{q}_1, \boldsymbol{b}_3 = \boldsymbol{q}_1 - \boldsymbol{q}_2$, where $\boldsymbol{q}_{j+1} = R\left(\frac{2\pi}{3}\right) \boldsymbol{q}_j$. The hBN alignment generates a moiré potential acting only on the bottom graphene layer[78,82,83]

$$V_{\text{moiré}}(\boldsymbol{r}) = V_1 e^{i\psi} \sum_{j=1}^{3} e^{i\boldsymbol{b}_j \cdot \boldsymbol{r}} \begin{pmatrix} 1 & \omega^{-j} \\ \omega^{j+1} & \omega \end{pmatrix} + h.c.$$

where $\omega = \exp\left(\frac{2\pi i}{3}\right)$, and $V_1$, $\psi$ are moiré coupling parameters. Note that we neglect the spatially uniform $V_0$ component of the hBN-induced potential for simplicity, and we use $\psi = 16.55°$. The Hamiltonian for valley $K'$ can be obtained by time reversal symmetry.

For $\nu = +1$, the moiré-distant regime in these conventions corresponds to $V > 0$.

### B. Interactions

We consider dual-gate screened Coulomb interactions $V(\boldsymbol{q}) = \frac{e^2}{2\epsilon_0 \epsilon_r q} \tanh(q d_{sc})$, where $d_{sc} = 10$ nm is the gate distance, and $\epsilon_r$ is the relative permittivity. Note that the interacting continuum model includes both spins as well as both valleys. Specifying the interaction term requires selecting an 'interaction scheme'. Intuitively, this is equivalent to choosing a reference density from which interactions are measured from. A detailed discussion of the interaction scheme can be found in refs.79,84. In this work, we consider the charge-neutrality (CN) and average (AVE) interaction schemes. In the CN scheme, the reference density corresponds to occupying all of the moiré valence bands. In the AVE scheme, the reference density corresponds to a uniform homogeneous background at neutrality.

### C. Sign of the Chern number in v = 1 Hartree-Fock calculations

Prior Hartree-Fock (HF) studies of RnG/hBN at $v = +1$ in the moiré distant regime ($V > 0$) have found spin-valley polarized insulators[79,81,85-87]. We are interested in the Chern numbers of these candidate states. In particular, we are interested in the valley Chern number, which can be extracted via the Chern number of a state polarized in valley $K$. Therefore, we will focus on $K$-polarized phases in the following. We define the Chern number of a band with Bloch function $|n(\mathbf{k})\rangle$ as

$$C = \frac{1}{2\pi} \int d\mathbf{k} \Omega(\mathbf{k})$$
$$\Omega(\mathbf{k}) = i \left( \langle \partial_{k_x} n | \partial_{k_y} n \rangle - \langle \partial_{k_y} n | \partial_{k_x} n \rangle \right)$$

where $\Omega(\mathbf{k})$ is the Berry curvature. We note that the conduction band minimum in the $K$ valley of RnG has $\Omega(\mathbf{k}) > 0$ for $V > 0$. In these conventions, the Streda's formula reads[77,88]

$$\frac{\delta n_e}{\delta B_\perp} = \frac{e}{2\pi \hbar} C,$$

where the electron charge is $-e < 0$. Therefore, an electronic band with $C < 0$ loses states in a positive perpendicular magnetic field $B_\perp > 0$.

We perform self-consistent HF calculations of $v = +1$ RHG/hBN ($n = 6$), projecting into the lowest five moiré conduction bands per spin and valley. In Extended Data Fig. 11, we present the phase diagram for the Chern number for a range of interlayer potentials $V$, moiré potential strengths $V_1$, twist angles $\theta$, and interaction schemes. For each parameter, we repeat the HF calculation for at least 15 initial seeds, and consider the Chern numbers (assuming polarization in valley $K$) of all gapped solutions, regardless of whether they are the lowest energy solution.

As shown in Extended Data Fig. 11, we find a competition between $C = 0, 1$ states, which has been reported previously[79,81,85-87]. We find that the calculations for the smaller twist angle ($\theta = 0.2°$) tend to relatively favor $C = 0$ compared to calculations at the larger twist angle ($\theta = 0.5°$). This trend versus twist angle has also been reported in previous theoretical studies[79,81,85-87]. We caution that mean-field theory may not correctly determine the phase competition as proposed in refs.89-91. However, we anticipate that HF is at least able to capture the set of candidate low-energy phases. In our HF calculations though, we are not able to find any $C = -1$ solutions. We note that previous exact diagonalization calculations on rhombohedral pentalayer graphene aligned with hBN also do no find any $C = -1$ states[79,84]. This strongly suggests that valley $K$ does not admit a candidate $C = -1$ phase, at least within existing continuum models.

### D. Orbital magnetization

We compute the orbital magnetization of the spin-valley polarized states in RHG/hBN. Consider first a general non-interacting Bloch Hamiltonian $H(\mathbf{k})$ with Bloch functions $|n(\mathbf{k})\rangle$ and energies $\epsilon_n(\mathbf{k})$, where $n$ indexes the bands (which includes spin and valley indices as well). The expression for the orbital magnetization is[77,88,92-94]

$$M_z = \frac{e}{\hbar} Im \int \frac{d\mathbf{k}}{(2\pi)^2} \sum_n \left\langle \partial_{k_x} n | H(\mathbf{k}) + \epsilon_n(\mathbf{k}) - 2\mu | \partial_{k_y} n \right\rangle \theta(\mu - \epsilon_n(\mathbf{k})),$$

where $\mu$ is the chemical potential, and the Heaviside function $\theta(\mu - \epsilon_n(\mathbf{k}))$ restricts the summation/integral to occupied states. $M_z$ can be decomposed into self-rotation and Chern magnetic moments

$$m_n^{SR}(\mathbf{k}) = \frac{e}{\hbar} Im \left\langle \partial_{k_x} n | (H(\mathbf{k}) - \epsilon_n(\mathbf{k})) | \partial_{k_y} n \right\rangle$$

$$= -\frac{e}{\hbar} Im \sum_{m \neq n} \frac{\langle n | \partial_{k_x} H(\mathbf{k}) | m \rangle \langle m | \partial_{k_y} H(\mathbf{k}) | n \rangle}{\epsilon_n(\mathbf{k}) - \epsilon_m(\mathbf{k})}$$

$$m_n^C(\mathbf{k}) = -\frac{2e}{\hbar} Im \left\langle \partial_{k_x} n | (\mu - \epsilon_n(\mathbf{k})) | \partial_{k_y} n \right\rangle$$

$$= -\frac{2e}{\hbar} Im \sum_{m \neq n} \frac{\langle n | \partial_{k_x} H(\mathbf{k}) | m \rangle \langle m | \partial_{k_y} H(\mathbf{k}) | n \rangle}{(\epsilon_n(\mathbf{k}) - \epsilon_m(\mathbf{k}))^{\wedge}2} [\mu - \epsilon_n(\mathbf{k})]$$

$$= \frac{e}{\hbar} \Omega_n(\mathbf{k})(\mu - \epsilon_n(\mathbf{k}))$$

Hence within the gap of an insulating state, the orbital magnetization changes proportionally to the total Chern number of the filled bands as the chemical potential is tuned. Note that for numerical computations, we use the second equations for each of $m^C$ and $m^{SR}$, since derivatives of the Bloch Hamiltonian do not require gauge-fixing when expressed in the plane wave basis. We approximate the derivatives by finite differences on the momentum mesh.

Applying the above formalism to HF calculations is associated with some caveats. First, we assume that the same equations for orbital magnetization can be directly used just by replacing $H(\mathbf{k}), |n(\mathbf{k})\rangle, \epsilon_n(\mathbf{k})$ with the HF Hamiltonian, HF Bloch functions, and HF energies respectively. Second, our HF calculations are performed within a projected set of active bands, while the orbital magnetization formula requires information about all bands. Therefore, after the HF calculation, we embed the projected HF density matrix $P_{proj.}$ into the full unprojected Hilbert space by assuming all remote valence (conduction) bands are fully filled (empty), leading to an unprojected density matrix $P_{unproj.}$. We then choose $H(\mathbf{k})$, and its associated Bloch functions and energies, to correspond to the one-shot HF Hamiltonian corresponding to $P_{unproj.}$. Note that $P_{unproj.}$ Is generally not a self-consistent HF solution of the unprojected Hamiltonian.

As shown in Extended Data Fig. 12, we find that the orbital magnetization $M_z$ for the $C = 1$ HF state polarized in valley $K$ is always negative for both interaction schemes, both twist angles $\theta = 0.2°, 0.5°$, and both interaction strengths $\epsilon_r = 5,10$. We note that $M_z$ remains negative across the whole range of chemical potentials within the HF gap. We find that stronger interactions and smaller $V$ lead to larger absolute values $|M_z|$, which remains above $5\mu_B$ for the parameters considered here.

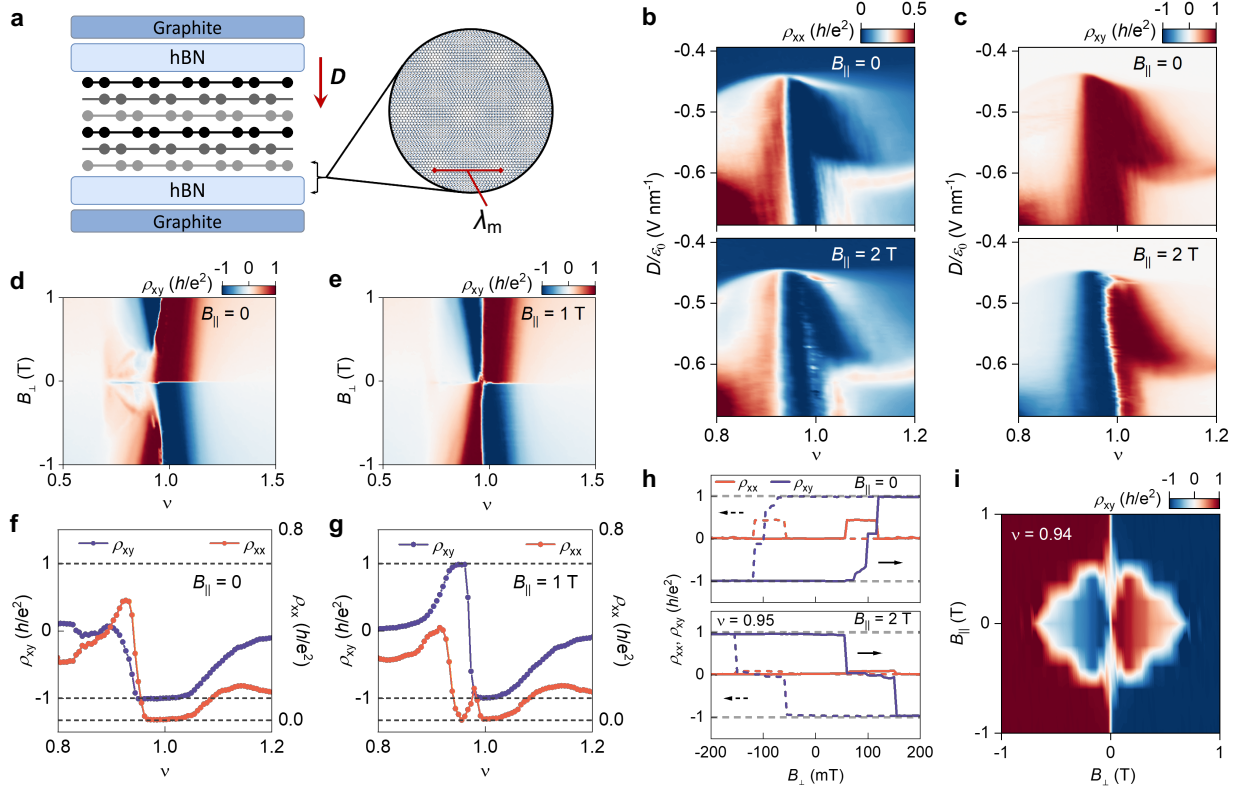

**Fig.1|Chirality switching of the CI at ν = 1. a,** Schematic of the device configuration. The illustration shows the moiré interface between RHG and bottom hBN. $\lambda_m$ represents moiré periodicity. **b,c,** Phase diagrams of longitudinal resistivity $\rho_{xx}$ (**b**) and Hall resistivity $\rho_{xy}$ (**c**) as a function of moiré filling factor ν and electric displacement field $D/\varepsilon_0$ measured at $B_\parallel = 0$ (top panel) and $B_\parallel = 2$ T (bottom panel). The Chern insulator state at ν = 1 is split into two regions with $C = -1$ and $C = 1$ under a finite $B_\parallel$. For the maps at $B_\parallel = 0$ and 2 T, an additional $B_\perp = 40$ mT and 100 mT was applied, respectively. **d,e,** Landau fan diagrams of $\rho_{xy}$ measured at $B_\parallel = 0$ (**d**) and $B_\parallel = 1$ T (**e**), respectively, with fixed $D/\varepsilon_0 = -0.53$ V nm$^{-1}$. **f,g,** The $\rho_{xx}$ and $\rho_{xy}$ versus ν obtained at $B_\parallel = 0$ (**f**) and $B_\parallel = 1$ T (**g**), respectively, with fixed $D/\varepsilon_0 = -0.53$ V nm$^{-1}$ and $B_\perp = -0.1$ T. **h,** The IQAHE at $B_\parallel = 0$ (top panel) and $B_\parallel = 2$ T (bottom panel), with fixed $D/\varepsilon_0 = -0.58$ V nm$^{-1}$ and ν = 0.95. The IQAHE is stabilized at $B_\parallel = 2$ T, but with an opposite chirality compared to the case with $B_\parallel = 0$. The data of $\rho_{xx}$ and $\rho_{xy}$ have been symmetrized and antisymmetrized, respectively. **i,** Phase diagram of antisymmetrized $\rho_{xy}$ versus $B_\perp$ and $B_\parallel$ measured at fixed ν = 0.94 and $D/\varepsilon_0 = -0.53$ V nm$^{-1}$. The data for negative $B_\parallel$ are obtained by symmetrizing the data from positive $B_\parallel$, as the phase diagram exhibits symmetry with respect to the direction of the $B_\parallel$, evidenced by the similar phase diagram obtained from device 3 (Extended Data Fig. 3e). Note that only data shown in **h** are measured from device 2, all other data in Fig.1 are taken from device 1.

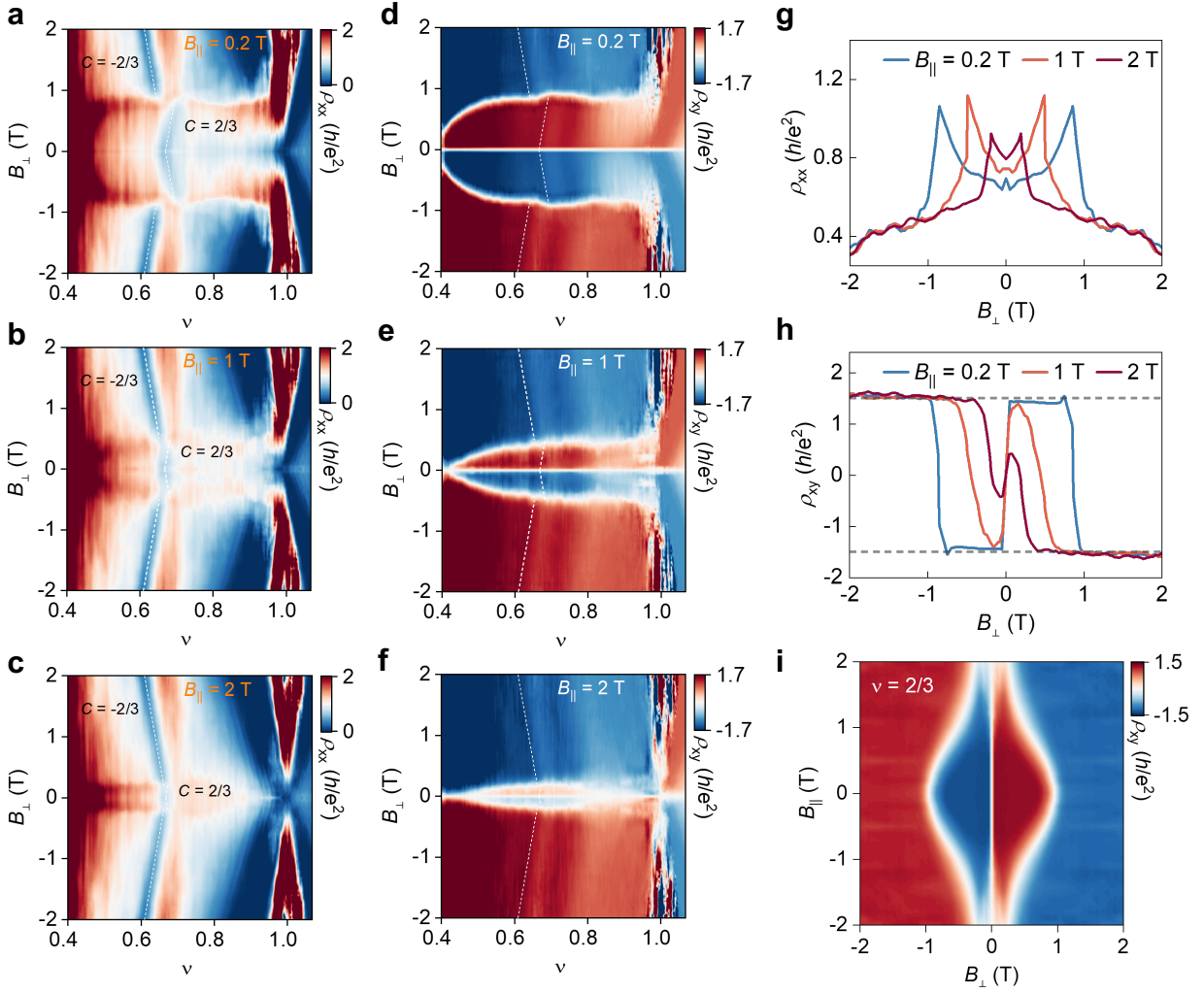

**Fig.2|Topological phase transitions of the ν = 2/3 FCI. a-c,** Landau fan diagrams of symmetrized $\rho_{xx}$ as a function of ν and $B_\perp$ measured at fixed $D/\varepsilon_0$ = -0.74 V nm$^{-1}$ and different in-plane magnetic fields ($B_\parallel$ = 0.2 T (**a**), 1 T (**b**), 2 T (**c**)). White dashed lines are guides to the eye for trajectories of ν = 2/3 state with a $C = \pm 2/3$ gap. **d-f,** Phase diagrams of antisymmetrized $\rho_{xy}$ versus ν and $B_\perp$ measured at fixed $D/\varepsilon_0$ = -0.74 V nm$^{-1}$ and different $B_\parallel$. The dashed lines are assigned based on the position of dips shown in the $\rho_{xx}$ maps under the same $B_\parallel$ in **a-c**. **g,h,** Line cuts of symmetrized $\rho_{xx}$ (**g**) and antisymmetrized $\rho_{xy}$ (**h**) versus $B_\perp$ measured at different $B_\parallel$, along the dashed lines from **a-f**, respectively. With $B_\parallel$ increasing, the $C = -2/3$ FCI state extends toward low $B_\perp$, while the $C = 2/3$ state is gradually suppressed. **i,** Phase diagrams of antisymmetrized $\rho_{xy}$ as a function of $B_\perp$ and $B_\parallel$ measured at fixed ν = 2/3 and $D/\varepsilon_0$ = -0.74 V nm$^{-1}$. The data for negative $B_\parallel$ are obtained by symmetrizing the data from positive $B_\parallel$, as the phase diagram exhibits symmetry with respect to the direction of the $B_\parallel$, evidenced by the similar phase diagram obtained from device 3 (Extended Data Fig. 3f).

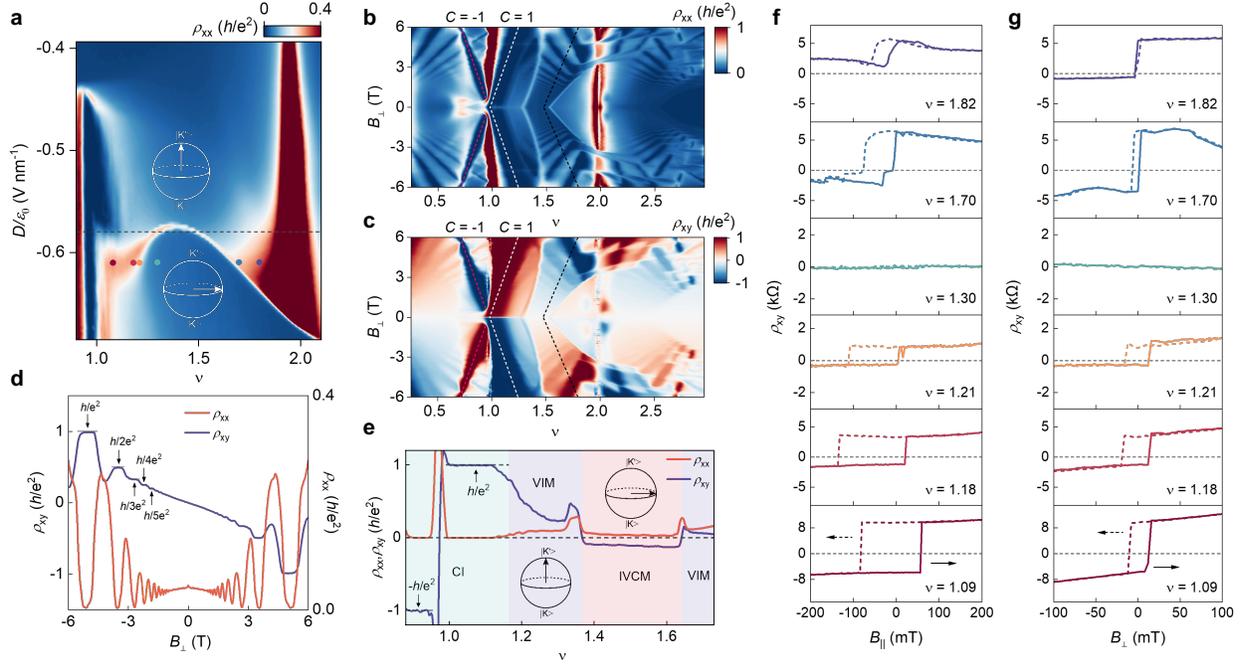

**Fig.3 | Phase transitions, intervalley coherence and AHE at 1 < ν < 2. a,** Phase diagram of $\rho_{xx}$ versus ν and $D/\varepsilon_0$ obtained at zero magnetic field, demonstrating a phase boundary at 1 < ν < 2. **b,c,** Landau fan diagrams of symmetrized $\rho_{xx}$ (**b**) and antisymmetrized $\rho_{xy}$ (**c**) versus ν and $B_\perp$ measured at $D/\varepsilon_0$ = -0.58 V nm$^{-1}$ (corresponding to the dashed line in **a**). The white dashed lines correspond to the CI state with $C$ = 1, whereas the red dashed lines correspond to the CI state with $C$ = -1. **d,** Symmetrized $\rho_{xx}$ and antisymmetrized $\rho_{xy}$ versus $B_\perp$, featuring a series of singly degenerate quantum Hall states with $\rho_{xy} = h/pe^2$, where $p$ is an integer. The curves are extracted along the black dashed lines from **b** and **c**, respectively. **e,** Line cuts of symmetrized $\rho_{xx}$ and antisymmetrized $\rho_{xy}$ versus ν obtained at $B_\perp$ = 1 T and $D/\varepsilon_0$ = -0.58 V nm$^{-1}$, showing different phases including Chern insulator (CI), valley-imbalanced metal (VIM) and intervalley-coherent metal (IVCM). The inserted schematic illustrates the valley polarization of VIM and IVCM phases on the Bloch sphere. For ICVM, the valley polarization is in the in-plane direction. **f,g,** Magnetic hysteresis loops of $\rho_{xy}$ measured at $D/\varepsilon_0$ = -0.58 V nm$^{-1}$ when sweeping $B_\parallel$ (**f**) and $B_\perp$ (**g**) back and forth. Both of the $\rho_{xy}$ scans exhibit clear magnetic hysteresis windows at the phase boundary and VIM region, which vanishes within the IVCM region. Different colors represent the data obtained at different moiré filling factors, corresponding to the colored points in **a**. The dashed line corresponds to $\rho_{xy}$ = 0.

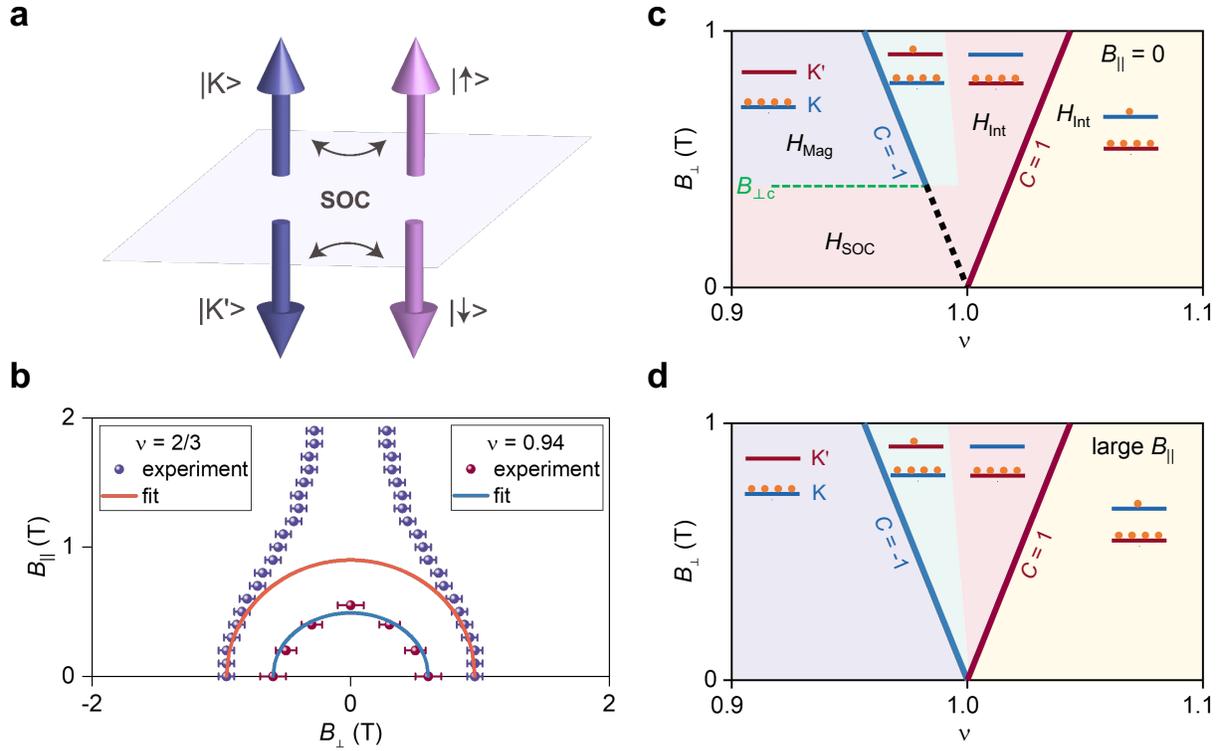

**Fig.4 | The role of spin-orbit coupling effect. a,** Illustration of the effective SOC, which couples the spin and valley degrees of freedom. **b,** The magnetic field phase boundary for chirality reversal. The experimental data points are inferred from the phase boundaries in Fig. 1i and 2i, with error bars representing the maximum range of $B_\perp$ with Hall resistivity sign reversal. The fitting curves are semi-ellipses, given by $\frac{B_\perp^2}{a^2} + \frac{B_\parallel^2}{b^2} = 1$, where $(a, b) = (0.6\ \text{T}, 0.49\ \text{T})$ for $\nu = 0.94$ and $(0.96\ \text{T}, 0.9\ \text{T})$ for $\nu = 2/3$. SOC strength $\lambda_I = 40\ \mu\text{eV}$ is used in the fit. The $D/\varepsilon_0$ is fixed at $-0.53\ \text{V nm}^{-1}$ for $\nu = 0.94$ and $-0.74\ \text{V nm}^{-1}$ for $\nu = 2/3$. **c,d,** Schematic phase diagram of valley polarization near $\nu = 1$, for $B_\parallel = 0$ (**c**) and large $B_\parallel$ (**d**) respectively. The regions filled with different colors represent distinct valley-polarized electronic states. Specifically, on the left of the blue (red) solid line, which corresponds to $C = -1$ ($C = +1$), electrons fully occupy K (K') valley. We labeled different regions in **c** according to the dominant terms in the Hamiltonian: $H_{\text{Mag}}$ indicates spin and valley magnetization dominance, $H_{\text{SOC}}$ corresponds to SOC dominance, and $H_{\text{Int}}$ denotes interaction-driven regimes. See Methods for more details.

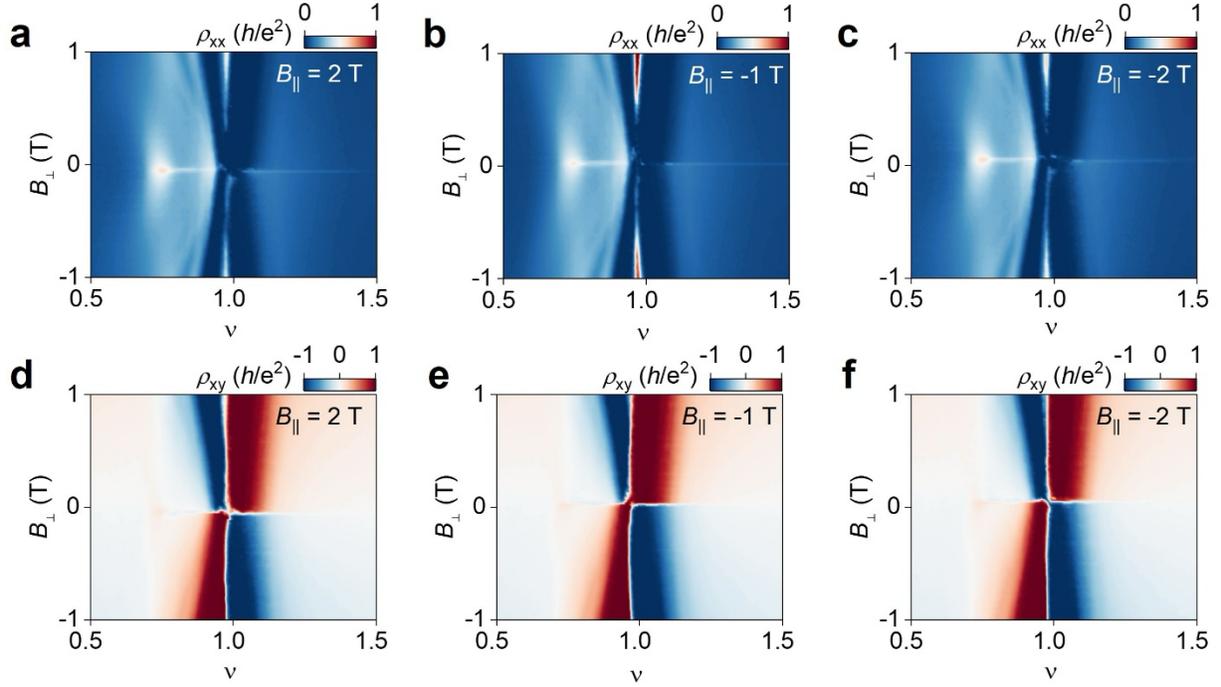

**Extended Data Fig.1|Landau fan diagrams for ν = 1 state at different $B_\parallel$. a-c,** Landau fan diagrams of $\rho_{xx}$ measured at $B_\parallel$ = 2 T (**a**), -1 T (**b**) and 2 T (**c**). **d-f,** Landau fan diagrams of $\rho_{xy}$ measured at $B_\parallel$ = 2 T (**d**), -1 T (**e**) and 2 T (**f**). Consistent with Fig. 1, the $C$ = -1 state can extend to $B_\perp$ = 0 under the influence of $B_\parallel$.

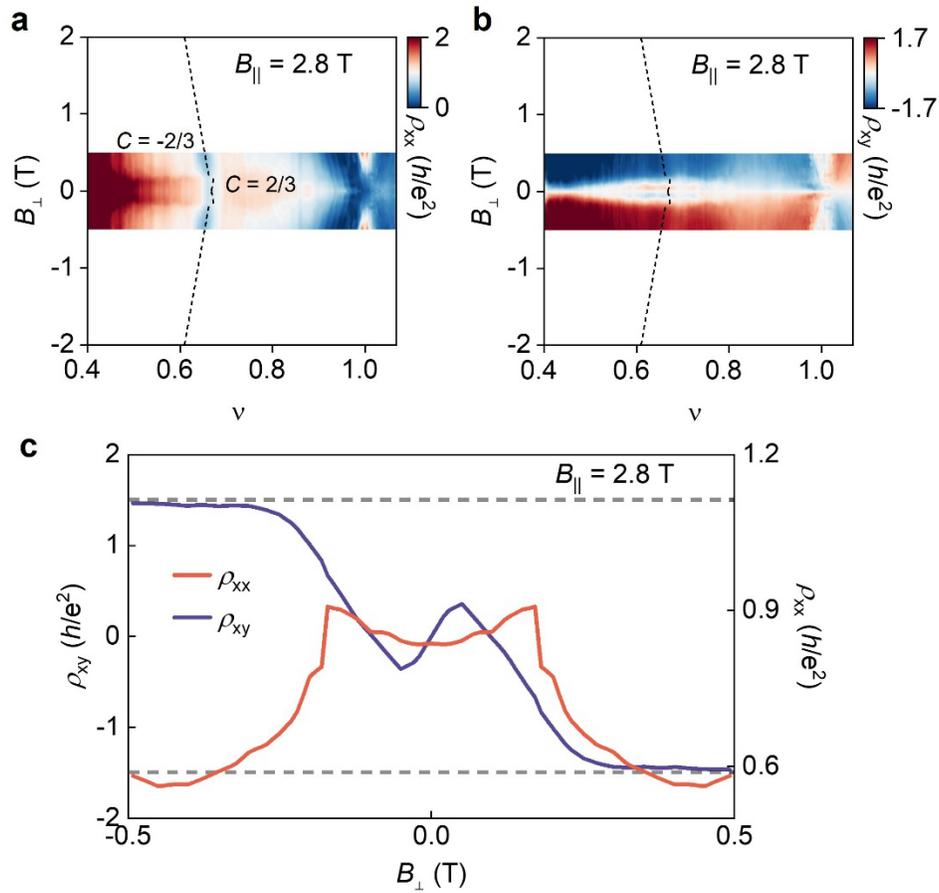

**Extended Data Fig.2 | Competing FCI states at large $B_∥$. a,b,** phase diagrams of symmetrized $ρ_{xx}$ (**a**) and antisymmetrized $ρ_{xy}$ (**b**) versus ν and $B_⊥$ measured at $D/ε_0$ = -0.74 V nm$^{-1}$ and $B_∥$ = 2.8 T. **c,** Line cuts of symmetrized $ρ_{xx}$ and antisymmetrized $ρ_{xy}$ versus $B_⊥$ corresponding to the dashed lines in **a** and **b**. The second Hall resistivity sign reversal still persists even as $B_∥$ increases to 2.8 T.

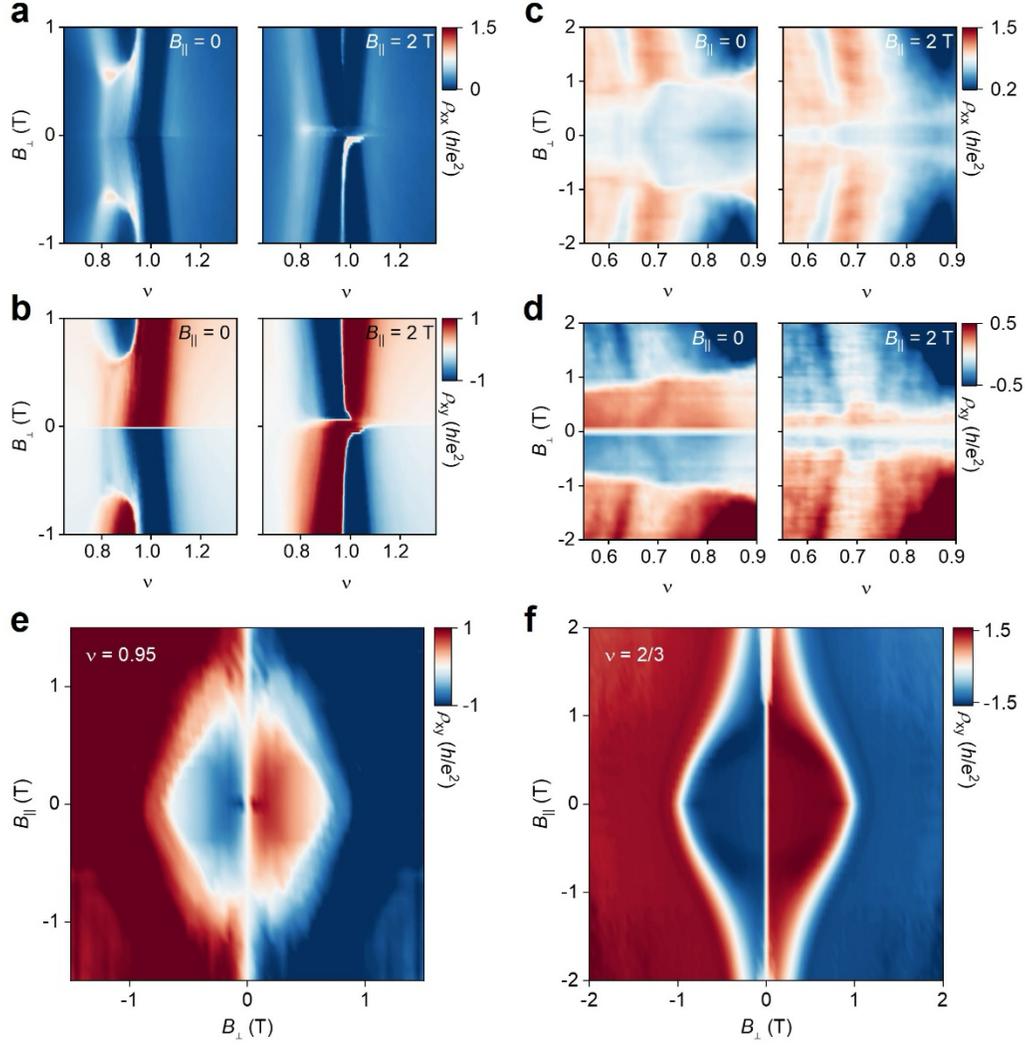

**Extended Data Fig.3|Topological phase transitions at both integer and fractional states observed in device 2 and device 3**. **a,b,** Landau fan diagrams of $\rho_{xx}$ (**a**) and $\rho_{xy}$ (**b**) measured at $B_\parallel = 0$ (left) and $B_\parallel = 2$ T (right), respectively. The $D/\varepsilon_0$ is fixed at -0.58 V nm$^{-1}$. **c,d,** Phase diagrams of symmetrized $\rho_{xx}$ (**c**) and antisymmetrized $\rho_{xy}$ (**d**) versus ν and $B_\perp$ measured at $B_\parallel = 0$ (left) and $B_\parallel = 2$ T (right), respectively, with fixed $D/\varepsilon_0 = -0.81$ V nm$^{-1}$. Note that the data of **a-d** are obtained from device 2. **e,f,** Phase diagrams of antisymmetrized $\rho_{xy}$ versus $B_\perp$ and $B_\parallel$ measured at ν = 0.95 (**e**) and ν = 2/3 (**f**) obtained from device 3, showing the similar phase boundary of chirality reversal to device 1. The $D/\varepsilon_0$ is fixed at 0.635 V nm$^{-1}$ (**e**) and 0.78 V nm$^{-1}$ (**f**), respectively.

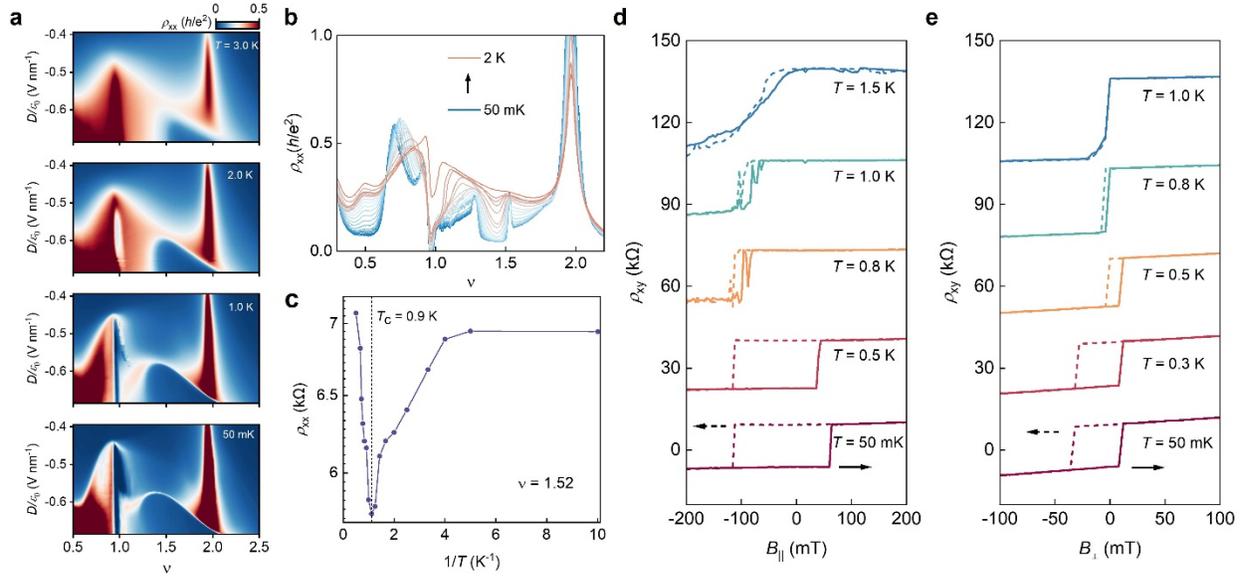

**Extended Data Fig.4 | Temperature dependence of phase transitions and AHE at 1 < ν < 2. a,** $\rho_{xx}$ maps of zero magnetic field measured at different temperatures, exhibiting the gradual merging process of the phase boundary and the $C = 1$ Chern gap. **b,c,** The weak insulating behavior of the phase boundary. The $\rho_{xx}$ versus ν measured at $D/\varepsilon_0 = $ -0.58 V nm$^{-1}$ (**b**) and the gradient colors represent a gradual increase in temperature. The $\rho_{xx}$ as a function of $1/T$ at ν = 1.52 and $D/\varepsilon_0 = $ -0.58 V nm$^{-1}$ (**c**). **d,e,** Temperature dependence of magnetic hysteresis loops at ν = 1.52 and $D/\varepsilon_0 = $ -0.58 V nm$^{-1}$, when sweeping $B_\parallel$ (**d**) and $B_\perp$ (**e**) back and forth at different temperatures. Each sweep is offset by 30 kΩ for clarity.

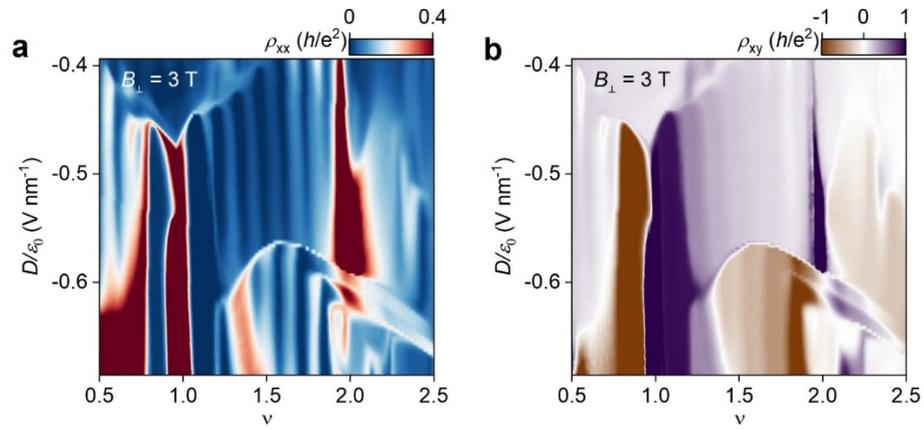

**Extended Data Fig.5|Quantum oscillations at high $B_\perp$. a,b,** Phase diagram of $\rho_{xx}$ (**a**) and $\rho_{xx}$ (**b**) versus $\nu$ and $D/\varepsilon_0$ measured at $B_\perp = 3$ T. Landau levels with a degeneracy of one are formed both above and below the VHS.

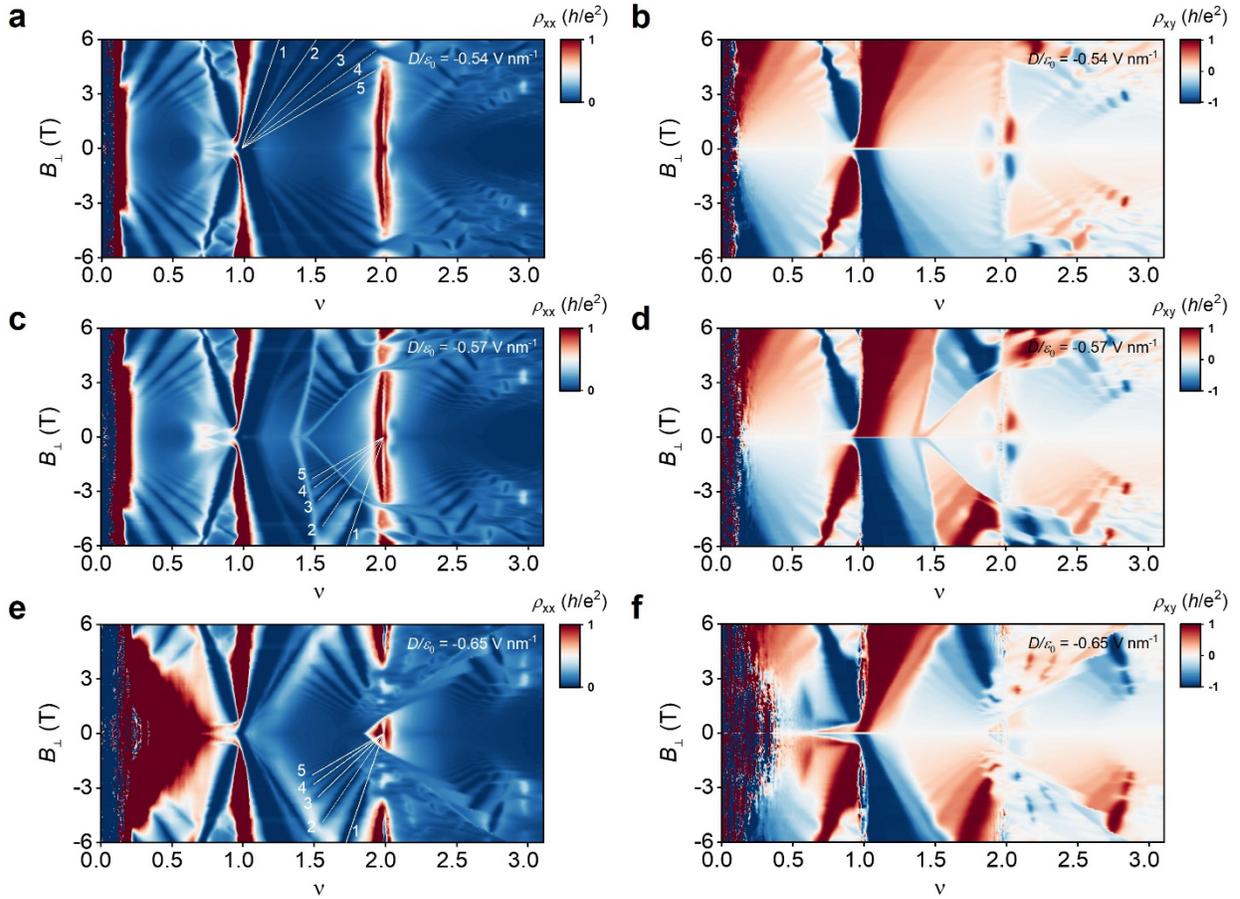

**Extended Data Fig.6 | Landau fan diagrams at different electric displacement fields. a,b,** Landau fan diagrams of symmetrized $\rho_{xx}$ (**a**) and antisymmetrized $\rho_{xy}$ (**b**) measured at $D/\varepsilon_0 = -0.54$ V nm$^{-1}$, where the phase boundary is absent and Landau levels appearing within $1 < \nu < 2$ are diverging from $\nu = 1$ state. The white lines serve as visual guides for the trajectories of the quantum Hall states. **c,d,** Landau fan diagrams of symmetrized $\rho_{xx}$ (**c**) and antisymmetrized $\rho_{xy}$ (**d**) obtained at the critical electric field ($D/\varepsilon_0 = -0.57$ V nm$^{-1}$) at which the phase boundary exactly appears. Landau levels inside the dome can be traced back to $\nu = 2$ state. **e,f,** Landau fan diagrams of symmetrized $\rho_{xx}$ (**e**) and antisymmetrized $\rho_{xy}$ (**f**) obtained at $D/\varepsilon_0 = -0.65$ V nm$^{-1}$. Landau levels originating from $\nu = 2$ state still persist.

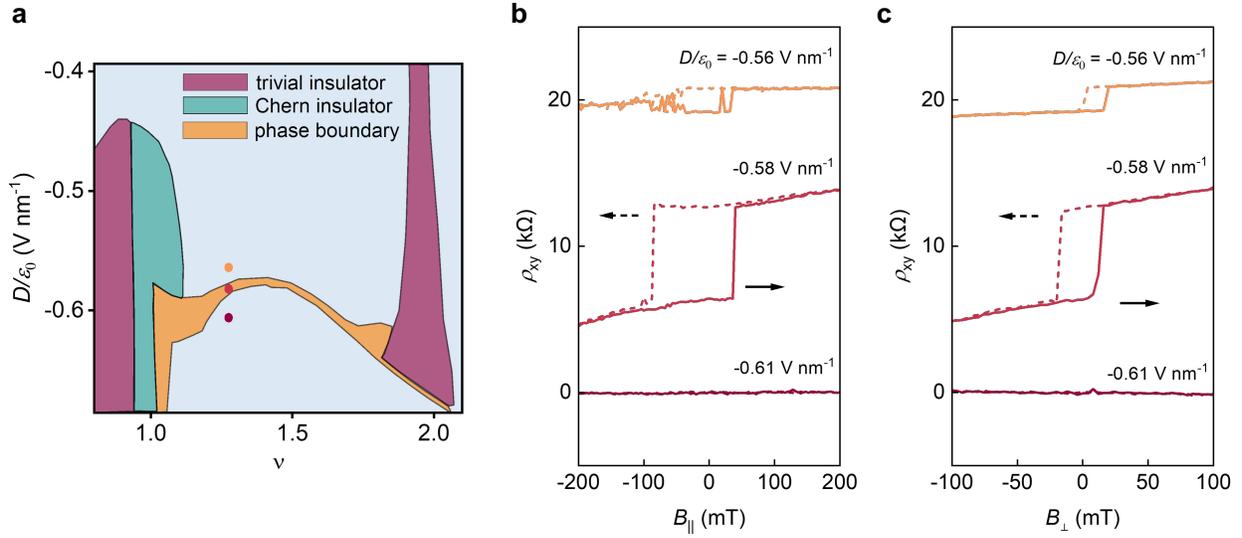

**Extended Data Fig.7 | Electric field dependence of the AHE at 1 < ν < 2. a,** Schematic of the phase diagram at 1 < ν < 2. **b,c,** Magnetic hysteresis loops of $\rho_{xy}$ measured at fixed ν = 1.27, with the *D* field corresponding to the positions marked with different colors in **a**. Each sweep is offset by 10 kΩ for clarity. Consistent with Fig. 3, the AHE occurs at the dome and gradually vanishes when moving below the phase boundary.

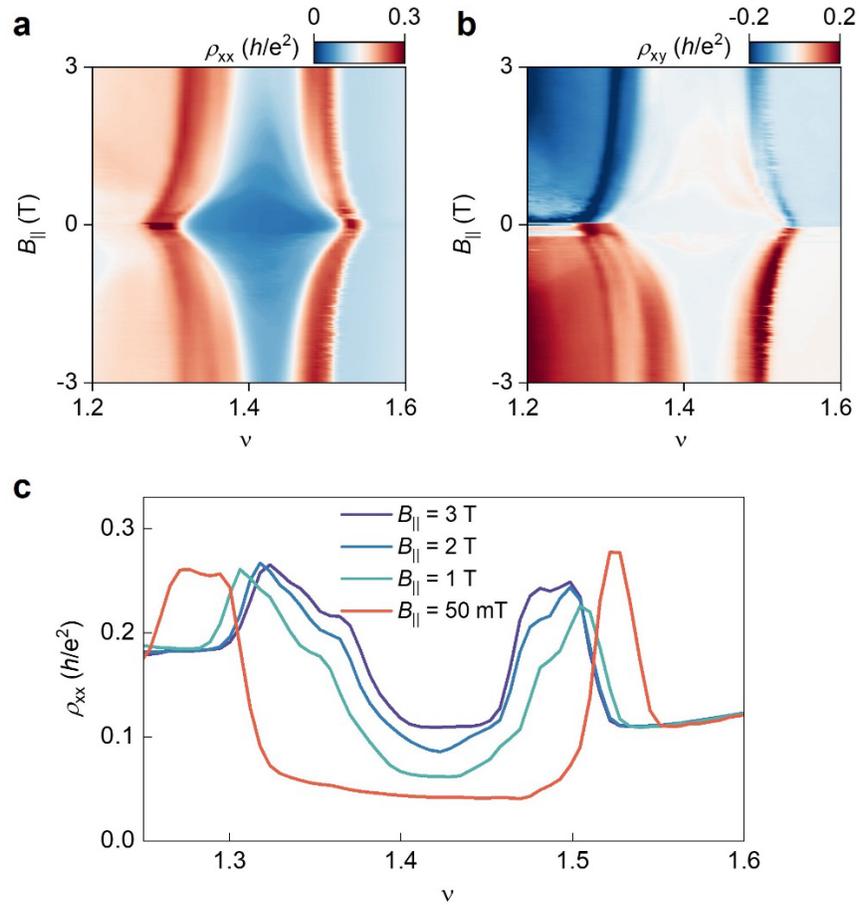

**Extended Data Fig.8 | $B_∥$-dependence of the phase transition at 1 < ν < 2. a,b,** phase diagrams of $ρ_{xx}$ (**a**) and $ρ_{xy}$ (**b**) as a function of ν and $B_∥$ measured at $D/ε_0 = -0.58$ V nm$^{-1}$. The phase boundary exhibits a nonlinear dependence on $B_∥$. **c,** The $ρ_{xx}$ versus ν measured at different $B_∥$. At low $B_∥$, the phase boundary is highly sensitive to $B_∥$, whereas at high $B_∥$, it becomes nearly independent of $B_∥$.

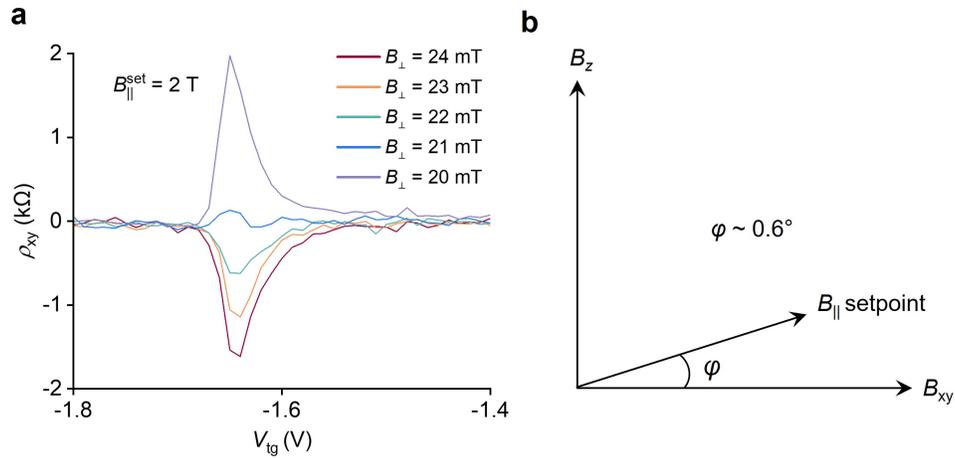

**Extended Data Fig.9|Estimation the perpendicular component of the in-plane magnetic field.**
**a,** Hall resistivity $\rho_{xy}$ as a function of top gate voltage, measured in the Bernal-stacked region of device 1. The bottom gate is fixed at 2.5 V. We set the in-plane magnetic field at 2 T and apply different perpendicular magnetic fields to counteract the perpendicular component of the in-plane field. The results indicate that the 2 T in-plane magnetic field has an approximately -21 mT perpendicular component. **b,** Estimation of the offset angle of the in-plane magnetic field direction. From **a**, we can obtain the offset angle $\varphi = \sin^{-1}(0.021/2) \approx 0.6°$.

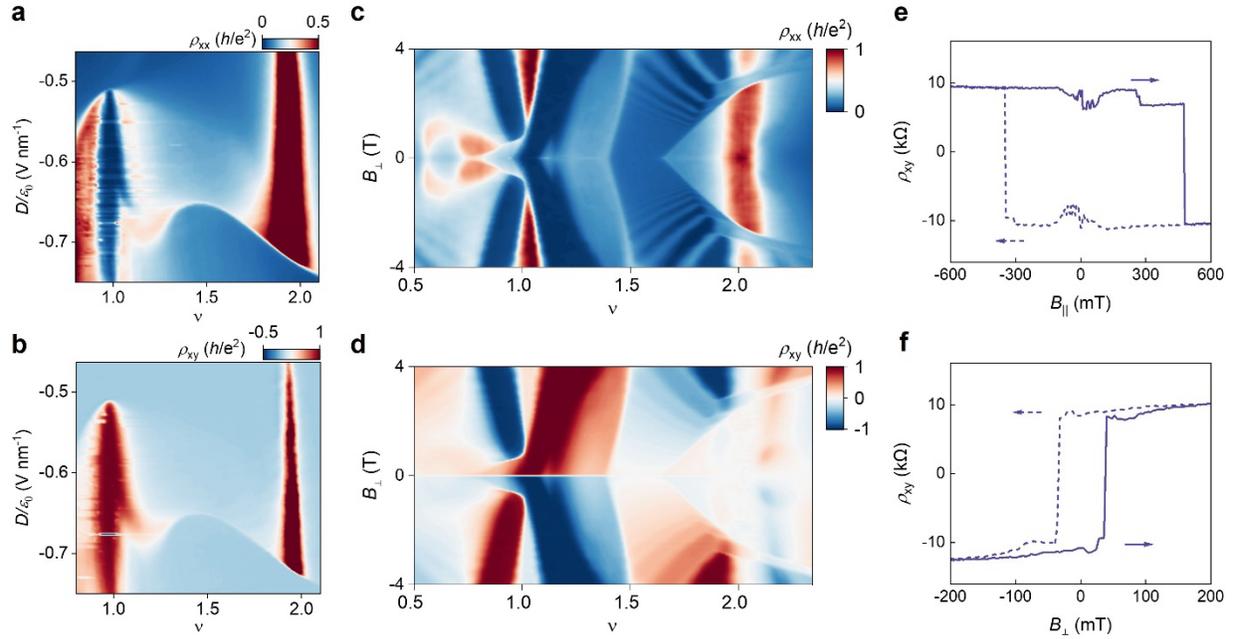

**Extended Data Fig.10 | Phase transitions with AHE observed in device 2. a,b,** Phase diagrams of $\rho_{xx}$ (**a**) and $\rho_{xy}$ (**b**) versus $\nu$ and $D/\varepsilon_0$ measured at zero magnetic field, showing similar phase transition features at $1 < \nu < 2$ to device 1. **c,d,** Landau fan diagrams of symmetrized $\rho_{xx}$ (**c**) and antisymmetrized $\rho_{xy}$ (**d**) obtained at $D/\varepsilon_0$ = -0.69 V nm$^{-1}$. **e,f,** Magnetic hysteresis loops of $\rho_{xy}$ measured at $\nu$ = 1.18 and $D/\varepsilon_0$ = -0.66 V nm$^{-1}$ when sweeping $B_{||}$ (**b**) and $B_\perp$ (**c**) back and forth.

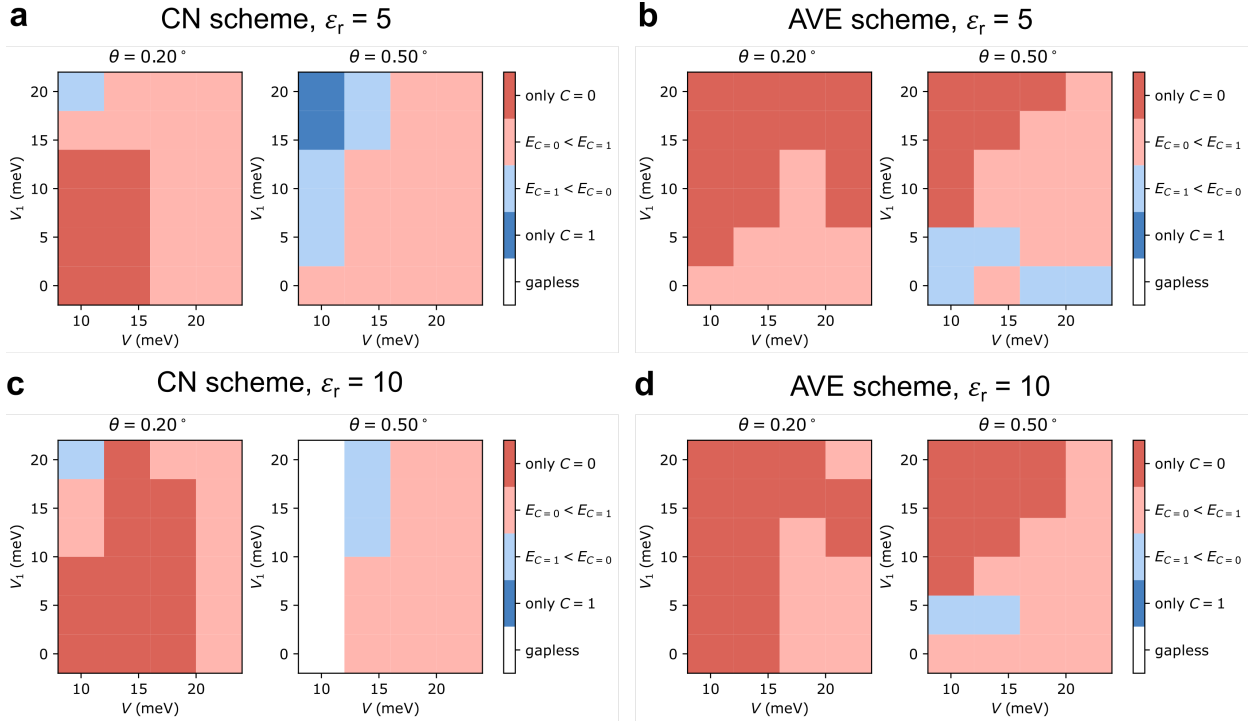

**Extended Data Fig.11|ν = 1 moiré-distant Hartree-Fock calculations for $\theta = 0.2°, 0.5°$ RHG/hBN with the CN scheme (a,c) and the AVE scheme (b,d).** For all parameters that are not gapless, the lowest HF solution is fully spin-valley polarized and gapped. We indicate the Chern numbers of the gapped solutions (which we take to be polarized in valley $K$ without loss of generality), and their relative energy ordering. The label 'only $C = 0$' means that all gapped solutions have $C = 0$, and analogously for 'only $C = 1$'. $E_{C=0} < E_{C=1}$ means that the $C = 0$ state is the HF ground state, but we are able to stabilize a (metastable) $C = 1$ solution. We do not find any solutions with $C = -1$. We consider relative permittivity $\epsilon_r = 5$ (**a,b**) and $\epsilon_r = 10$ (**c,d**). The system size is $12 \times 12$, and we project onto the five lowest moiré conduction bands. CN scheme represents the charge-neutrality interaction scheme. AVE scheme represents the average interaction scheme. See Methods for more details.

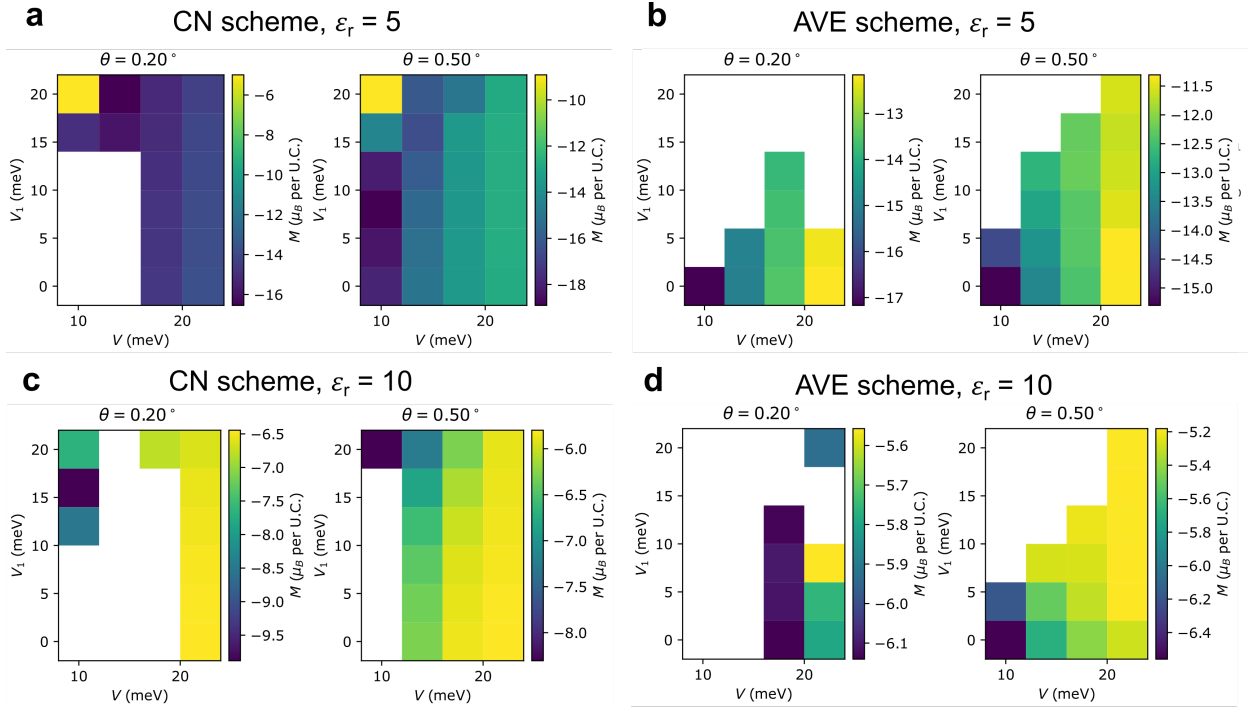

**Extended Data Fig.12 | ν = 1 moiré-distant Hartree-Fock calculations of the orbital magnetization $M_z$ of the $K$-valley $C = 1$ HF solution for $\theta = 0.2°, 0.5°$ RHG/hBN with the CN scheme (a,c) and the AVE scheme (b,d).** We consider relative permittivity $\epsilon_r = 5$ (**a,b**) and $\epsilon_r = 10$ (**c,d**). The chemical potential is set in the middle of the HF gap. White regions indicate parameters where we do not find a gapped $C = 1$ state. $M_z$ is plotted in units of $\mu_B = \frac{e\hbar}{2m_e}$ per moiré unit cell. The system size is $12 \times 12$, and we project onto the five lowest moiré conduction bands. CN scheme represents the charge-neutrality interaction scheme. AVE scheme represents the average interaction scheme. See Methods for more details.